\documentclass[sigconf]{acmart}

\AtBeginDocument{%
  }

\copyrightyear{2023}
\acmYear{2023}
\setcopyright{rightsretained}
\acmConference[MICRO '23]{56th Annual IEEE/ACM International Symposium on Microarchitecture}{October 28-November 1, 2023}{Toronto, ON, Canada}
\acmBooktitle{56th Annual IEEE/ACM International Symposium on Microarchitecture (MICRO '23), October 28-November 1, 2023, Toronto, ON, Canada}
\acmDOI{10.1145/3613424.3623786}
\acmISBN{979-8-4007-0329-4/23/10}

\usepackage{multirow, booktabs}
\usepackage{amsmath}
\usepackage{caption}
\usepackage{subcaption}
\usepackage{enumitem}
\usepackage{tikz}
\usepackage{yfonts}
\usepackage{dutchcal}
\usepackage{wrapfig}
\usepackage{floatflt}
\newcommand{\ie}{{\em i.e.}}
\newcommand{\eg}{{\em e.g.}}
\newcommand{\designName}{{HighLight}}

\definecolor{amber}{rgb}{1.0, 0.75, 0.0}

\usepackage{lipsum,xcolor}

\definecolor{SpringGreen}{RGB}{113, 181, 113}
\definecolor{LimeGreen}{RGB}{50, 205, 50}
\definecolor{BurntOrange}{RGB}{203,96,21}
\definecolor{Green}{RGB}{34,139,34}
\definecolor{BrickRed}{RGB}{170, 74, 68}
\definecolor{Orchid}{RGB}{229, 204, 255}

\usepackage{colortbl}
\begin{document}

\title{\designName: Efficient and Flexible DNN Acceleration with \\ Hierarchical Structured Sparsity } 

\author{Yannan Nellie Wu}
\affiliation{%
  \institution{MIT}
  \city{Cambridge}
  \state{MA}
  \country{USA}
}
\email{nelliewu@mit.edu}

\author{Po-An Tsai}
\affiliation{%
  \institution{NVIDIA}
  \streetaddress{}
  \city{Westford}
  \state{MA}
  \country{USA}}
\email{poant@nvidia.com}

\author{Saurav Muralidharan}
\affiliation{%
  \institution{NVIDIA}
  \city{Santa Clara}
  \state{CA}
  \country{USA}
}
\email{sauravm@nvidia.com}

\author{Angshuman Parashar}
\affiliation{%
  \institution{NVIDIA}
  \streetaddress{}
  \city{Westford}
  \state{MA}
  \country{USA}}
\email{aparashar@nvidia.com}

\author{Vivienne Sze}
\affiliation{%
  \institution{MIT}
  \city{Cambridge}
  \state{MA}
  \country{USA}
}
\email{sze@mit.edu}

\author{Joel S. Emer}
\affiliation{%
  \institution{MIT/NVIDIA}
  \streetaddress{}
  \city{Cambridge/Westford}
  \state{MA}
  \country{USA}}
\email{jsemer@mit.edu}

\renewcommand{\shortauthors}{ Wu et al.}

\begin{abstract}
Due to complex interactions among various deep neural network (DNN) optimization techniques, modern DNNs can have weights and activations that are dense or sparse with diverse sparsity degrees.  To offer a good trade-off between accuracy and hardware performance, an ideal DNN accelerator should have high flexibility to efficiently translate DNN sparsity into reductions in energy and/or latency without incurring significant complexity overhead.

This paper introduces hierarchical structured sparsity (HSS), with the key insight that we can systematically represent diverse sparsity degrees by having them hierarchically composed from multiple simple sparsity patterns. As a result, HSS simplifies the underlying hardware since it only needs to support simple sparsity patterns; this significantly reduces the sparsity acceleration overhead, which improves efficiency. Motivated by such opportunities, we propose a simultaneously efficient and flexible accelerator, named \designName, to accelerate DNNs that have diverse sparsity degrees (including dense). Due to the flexibility of HSS, different HSS patterns can be introduced to DNNs to meet different applications' accuracy requirements. Compared to existing works, \designName \ achieves a geomean of up to 6.4$\times$ better energy-delay product (EDP) across workloads with diverse sparsity degrees, and always sits on the EDP-accuracy Pareto frontier for representative DNNs.

\end{abstract}



\maketitle


\section{Introduction}
\label{sec:intro}

Modern deep neural networks (DNNs) can have weight and activation tensors with diverse discrete amounts of sparsity, \ie, \emph{sparsity degrees}, where sparsity is the percentage of zeros out of the total number of values in the tensor. This phenomenon is a result of complex interactions among various DNN optimization techniques in a large DNN model design space. For example, activations can be dense or sparse based on the choice of activation functions (\eg, ReLU~\cite{relu} introduces sparse activations, whereas Mish~\cite{mish} can result in much denser activations). Similarly, \emph{weight pruning} is often applied to over-parameterized DNN models, leading to zero-valued weights within the network.
Sparsity degrees for pruned DNNs vary depending on how amenable the given network is to sparsification (\eg, large models such as ResNet50~\cite{resnet} can sometimes be pruned to 80\% sparsity while still maintaining accuracy, while compact models such as EfficientNet~\cite{efficientnet} cannot be pruned as aggressively).

As a result, it is desirable to have a DNN accelerator that can translate any sparsity into efficiency, resulting in a good accuracy-efficiency trade-off. Specifically, the accelerator should be:
\begin{itemize}
    \item \textit{\textbf{Efficient}}: incurs low latency, energy, and area overhead cost, referred to as having \emph{low sparsity tax}, to implement the sparsity-related acceleration features. The sparsity tax can come from extra control logic, lack of data reuse, etc.
    \item \textit{\textbf{Flexible}}: supports \emph{diverse} sparsity degrees (including dense).
    "Support" refers to two capabilities: \textbf{(i)} process the DNN to produce functionally correct results; \textbf{(ii)} translate weight and activation sparsity into reductions in energy and/or latency.
\end{itemize}
Specifically, the accelerator has two goals: 
\begin{itemize}
\item for medium/high-sparsity DNNs, eliminate \emph{ineffectual operations} (\ie, compute and data movement involving zeros)~\cite{extensor} to introduce energy and/or latency savings; 
\item for low-sparsity DNNs, have similar energy efficiency and latency as a dense accelerator (\ie, have a low sparsity tax). 
\end{itemize}

\renewcommand{\arraystretch}{1.2}
\begin{table}[tb]
\resizebox{\columnwidth}{!}{
\centering
\begin{tabular}{c|c|c|c}
\textbf{\begin{tabular}[c]{@{}c@{}}Categories\end{tabular}}            & \textbf{\begin{tabular}[c]{@{}c@{}}Representative \\ Designs\end{tabular}} & \textbf{\begin{tabular}[c]{@{}c@{}}Sparsity \\ Tax\end{tabular}} & \textbf{\begin{tabular}[c]{@{}c@{}}Sparsity Degree \\ Diversity \end{tabular}} \\ \hline
Dense                                                                         
& TC~\cite{volta-white-paper}
& N/A    & \multicolumn{1}{c}{N/A}   \\ \hline

\multirow{2}{*}{\begin{tabular}[c]{@{}c@{}}Structured \\ Sparse\end{tabular}} 
& STC~\cite{ampere-white-paper}                                        
& \color{Green}{\textbf{Very Low}}  & \multicolumn{1}{c}{\color{BrickRed}{\textbf{Low}}}  \\ \cline{2-4} 

& S2TA~\cite{s2ta}                                                            
& \color{BurntOrange}{\textbf{Medium}}  & \multicolumn{1}{c}{\color{BurntOrange}{\textbf{Medium}}}   \\ \hline

\begin{tabular}[c]{@{}c@{}}Unstructured \\ Sparse\end{tabular}                
& DSTC~\cite{dstc}                                                          
& \color{BrickRed}{\textbf{High}}  & \multicolumn{1}{c}{\color{Green}{\textbf{Very High}}}                                                  \\ \hline
\textbf{\begin{tabular}[c]{@{}c@{}}HSS\end{tabular}}     & \textbf{\begin{tabular}[c]{@{}c@{}}Our Work\end{tabular}}             & \textbf{\color{LimeGreen}{\textbf{Low} }}                                                             & \multicolumn{1}{c}{\textbf{\textcolor{LimeGreen}{\textbf{High}}}}                                           \\ \bottomrule
\end{tabular}
}
\caption{\textbf{Comparison of designs from different DNN accelerator design categories. \emph{HSS} stands for hierarchical structured sparsity. An ideal design should have a low sparsity tax to achieve high efficiency and a very high number of supported sparsity degrees to achieve high flexibility.  }}
\label{tab:hss-existing-work}

\end{table}

\renewcommand{\arraystretch}{1.0}
However, to the best of our knowledge, none of the existing DNN accelerators achieve both goals~\cite{tpu-datacenter, samsung-npu, volta-white-paper, ampere-white-paper, s2ta, vector-sparse-tensor-core, dstc, eyeriss-v2, scnn, sparten, sparch, sigma}. Table~\ref{tab:hss-existing-work} describes the incurred sparsity tax and sparsity degree diversity for representative accelerators across different tensor accelerator categories.  \emph{Dense} accelerators~\cite{volta-white-paper, tpu-datacenter, diannao} have no sparsity tax, but never exploit sparsity. \emph{Structured sparse} accelerators~\cite{ampere-white-paper, s2ta, vector-sparse-tensor-core, structuredsparseFPGA} target DNNs whose sparsity is spatially constrained and introduce low-to-medium sparsity tax. However, they often only recognize a limited set of sparsity degrees. 
\emph{Unstructured sparse} DNN accelerators~\cite{dstc, sigma, outerspace, gamma} provide support for arbitrarily distributed zeros with diverse sparsity degrees, but pay a considerable sparsity tax (\eg, employ costly intersection units to locate nonzeros) for that flexibility. Thus, they are often inefficient for low-sparsity DNNs.
\emph{In short, the trend of DNNs containing tensors with diverse sparsity degrees challenges the fundamental design premise of many DNN accelerators.}

To address the limitations of existing work, we present a novel class of sparsity patterns named \emph{hierarchical structured sparsity} (HSS), with the insight that we can systematically represent diverse sparsity degrees by hierarchically composing them from simple sparsity patterns. Such simple sparsity patterns help us maintain a low sparsity tax by correspondingly simplifying the hardware that implements acceleration features for translating sparsity into reductions in energy/latency. Since there are many different ways of composing various simple sparsity patterns and designing their associated acceleration hardware, HSS opens up a promising design space. We evaluate the impact of various design decisions in such an HSS-based design space and propose an efficient and flexible accelerator, \designName\footnote{More information on \designName \ can be found at \url{http://emze.csail.mit.edu/highlight}}. 

\begin{figure}[t]%
    \centering
    \includegraphics[width=\columnwidth]{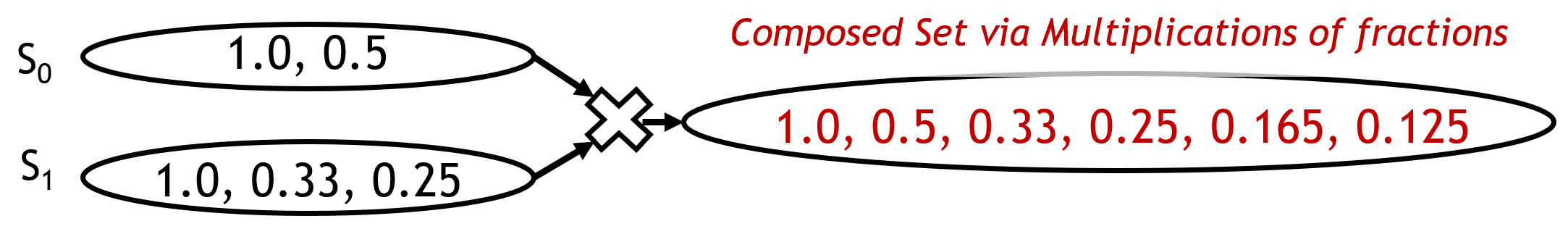} 
    \caption{\textbf{Composing two sets of density degrees, \emph{S$_0$} and \emph{S$_1$}, by multiplying the fractions in each set.} }
    \label{fig:mult_of_fractions}
    \vspace{5pt}
\end{figure}

The key insight of our design is that we leverage \emph{the properties of the multiplication of fractions} to: i) represent diverse structured sparsity degrees and ii) enable modularized low-sparsity-tax hardware support for each set of fractions to exploit the structured sparsity degrees. Fig.~\ref{fig:mult_of_fractions} illustrates the idea with two \emph{composable} sets of density degrees (where\textit{ density $=$ 1$-$sparsity}) represented as fractions. Composing the densities from \emph{S$_0$} and \emph{S$_1$} results in six density degrees. Thus, hardware with modularized support for each set naturally supports all six derived degrees.  


\vspace{2pt}
\noindent \emph{\textbf{This work makes the following key contributions:}}

\begin{enumerate}[topsep=0pt,itemsep=0pt,partopsep=0pt,parsep=0pt, leftmargin=12pt]
\item Proposes \emph{hierarchical structured sparsity (HSS)}, which allows a flexible representation of diverse sparsity degrees by hierarchically composing different sets of simple sparsity patterns. In addition, we propose a precise fibertree-based~\cite{dnn_syn_lec} sparsity specification to distinguish HSS from existing sparsity patterns. 

\item Proposes a simultaneously efficient and flexible hardware accelerator design, named \designName, that:
    \begin{itemize}[topsep=0pt,itemsep=0pt,partopsep=0pt,parsep=0pt, leftmargin=12pt]
        \item leverages modularity in HSS to enable modularized sparsity acceleration to exploit different sets of simple sparsity patterns in HSS at different architecture levels.
        \item introduces low-overhead sparsity acceleration hardware at each architecture level to provide efficient processing with a low sparsity tax.
    \end{itemize}

\item Demonstrates that DNNs can use HSS to meet various accuracy/efficiency requirements at various sparsity degrees.

\item To demonstrate efficiency and flexibility, \designName \ outperforms existing works with better overall hardware efficiency across workloads with diverse degrees in terms of both energy-delay-product (EDP) and energy-delay-squared (ED$^2$). 
\begin{itemize}[topsep=0pt,itemsep=0pt,partopsep=0pt,parsep=0pt, leftmargin=12pt]
    \item Compared to dense accelerators,  \designName\ achieves a geomean of 6.4$\times$ (and up to 20.4$\times$) lower EDP across DNN layers with diverse sparsity degrees (including dense) and is at EDP parity for dense DNN layers.
    \item Compared to sparse accelerators, \designName\ achieves a geomean of 2.7$\times$ (and up to 5.9$\times$) lower EDP and is at EDP parity for sparse DNN layers.
\end{itemize}


\end{enumerate}

\section{Background \& Motivation}

This section introduces the basics of sparse DNN acceleration, discusses the limitations of accelerators designed for different sparsity patterns (\ie, the distribution of zero and nonzero value locations), and motivates the need for a simultaneously efficient and flexible sparse DNN accelerator.

\subsection{Opportunities and Challenges}

The zero values in sparse DNNs can introduce a significant number of \emph{ineffectual computations}, whose results can be easily derived by applying the simple algebraic equalities of $X\times 0 = 0$ and $X + 0 = X$, without reading all the operands or doing the computations~\cite{extensor, micro-2022-sparseloop}. Thus, ineffectual computations introduce promising opportunities for accelerators to eliminate unnecessary hardware operations (\ie, buffer accesses and arithmetic calculations) and improve efficiency. 

However, to translate such opportunities into hardware savings, the accelerator faces the challenge of providing hardware support to identify nonzero values and evenly distribute them to parallel hardware components, referred to as \emph{workload balancing}. Workload balancing is important to ensure high utilization of the available resources, thus achieving maximum speedup. Often, the sparsity tax associated with such hardware support is highly related to the sparsity patterns that the accelerator aims to exploit.

\subsection{Limitations of Existing Accelerators}
\label{sec:sparsity-schemes}

In recent years, many sparse DNN accelerators
have been proposed to exploit ineffectual computations to reduce data movement and compute for different sparsity patterns
~\cite{dstc, scnn, eyeriss-v2, eyeriss-v1, sparten, GoSPA, ampere-white-paper, s2ta, gamma, vector-sparse-tensor-core, extensor, outerspace, sparch, sigma}.
At a high level, we can classify them into \emph{unstructured sparse accelerators} and \emph{structured sparse accelerators}. In the following sections, we will discuss their limitations both qualitatively and quantitatively.

\subsubsection{Unstructured Sparse Accelerators}
Unstructured sparse accelerators target DNNs with unstructured sparsity, which refers to an unconstrained distribution of zeros. Such patterns can be introduced to activations by activation functions or to weights by unstructured pruning that removes weights regardless of their locations in the tensor. 

Unstructured sparse accelerators have high flexibility to exploit arbitrarily distributed zeros with any sparsity degree. However, the hardware support for unstructured sparsity introduces a high sparsity tax since it cannot make any assumptions about the locations of nonzero values when trying to identify and distribute the effectual computations. Existing unstructured sparse accelerators either pay for expensive intersections to identify the effectual computations (\eg, SparTen~\cite{sparten} employs a prefix sum logic that occupies 55\% of its processing element area), or employ dataflows that identify effectual computations without intersections but require large, and thus expensive, accumulation buffers to hold the now randomly distributed output (\eg, the costly dataflow employed by DSTC~\cite{dstc}). Furthermore, since the number of effectual computations varies across sub-tensors within and across workloads, these accelerators can often only ensure perfect workload balance for a limited set of sparsity (\eg, DSTC~\cite{dstc} only ensures perfect workload balancing among columns of compute units when a sub-tensor's occupancy is a multiple of 32).

\noindent \textbf{Takeaway: unstructured sparse accelerators often support diverse sparsity degrees with a high sparsity tax.}

\subsubsection{Structured Sparse Accelerators}
Structured sparse accelerators target DNNs with structured sparsity, which refers to distributions of zeros with spatial constraints and is often introduced via \emph{structured pruning}~\cite{channel-pruning, NM-finetune, patdnn, s2ta}. Structured sparsity can have different spatial constraints for nonzero value locations. For example, one of the most popular structured sparsity patterns is the \emph{G:H sparsity pattern}, which mandates (at most) G elements to be nonzero within a block of H elements, and thus results in a density of G$/$H. For example, NVIDIA's Sparse Tensor Core (STC)~\cite{ampere-white-paper} employs a 2:4 pattern, which sparsifies two elements in every block of four elements~\cite{NM-finetune}, resulting in 50\% sparsity.

The predetermined constraints for nonzero value locations in structured sparsity make it much easier for hardware to identify the locations of the nonzeros and evenly distribute them to parallel hardware components (\eg, for G:H sparsity, the hardware can evenly assign G nonzeros to G compute units to balance the workload). As a result, accelerating a specific pattern is often efficient with a very low sparsity tax. However, existing structured sparse tensor accelerators often only accelerate  a very limited set of sparsity patterns (\ie, a few sparsity degrees). For example, the STC~\cite{ampere-white-paper} is only able to exploit the 2:4 pattern, whereas S2TA~\cite{s2ta} exploits a few $G:8$ patterns with design-specific constraints. 

\noindent \textbf{Takeaway: structured sparse accelerators often incur a low sparsity tax but only support a few sparsity degrees.}

\subsubsection{Quantitative Comparison}
To concretely demonstrate the limitations of each class of accelerators, without loss of generalizability, we quantitatively compare representative designs allocated with similar hardware resources: \textbf{(i) \emph{DSTC-like}~\cite{dstc}}: targets unstructured sparse DNNs with a high sparsity tax introduced by its costly dataflow; \textbf{(ii) \emph{STC-like}~\cite{ampere-white-paper}}: targets DNNs with weights that are dense or 2:4 sparse, introducing a low sparsity tax. 

\begin{figure}[t]%
    \centering
    \includegraphics[width=\columnwidth]{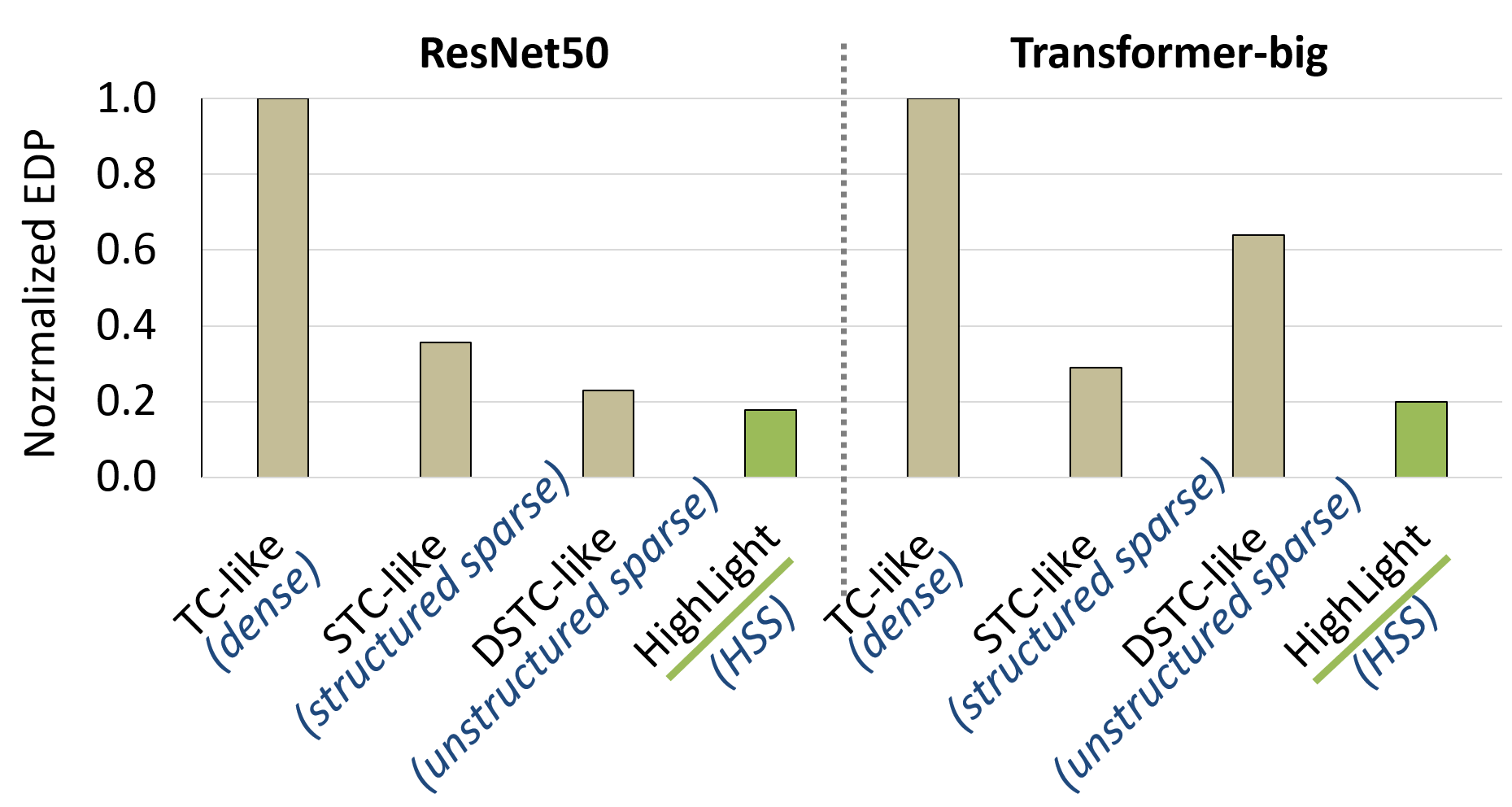} 
    \caption{\textbf{Normalized energy-delay product (EDP) of accelerators running two types of DNNs, pruned Transformer-Big~\cite{attention} and pruned ResNet50~\cite{resnet}. While ensuring similar accuracy (within 0.5\% difference), the DNNs were structured pruned for STC~\cite{ampere-white-paper} and \designName\ (our work) and unstructured pruned for DSTC. For both models, \designName \ achieves the lowest EDP while ensuring similar accuracy.} }
    \label{fig:hss_mov}
 
\end{figure}

To compare the designs, we normalize their energy-delay-product (EDP) to a dense accelerator, \emph{TC-like}~\cite{volta-white-paper}. In particular, we evaluate their EDP running two different DNN architectures, \emph{Transformer-Big}~\cite{attention} and \emph{ResNet50}~\cite{resnet}. For each DNN architecture, while ensuring similar accuracy (which we define as $<$0.5\% difference), \emph{TC-like} runs the dense version of the model, \emph{STC-like} runs a structured pruned version, and \emph{DSTC-like} runs an unstructured pruned version. Fig.~\ref{fig:hss_mov} shows the EDP of the three designs running all the GEMM layers in the DNNs.

\textbf{Inflexibility of Structured Sparse Designs:} As shown in Fig.~\ref{fig:hss_mov}, \emph{STC-like} is outperformed by \emph{DSTC-like} when running \emph{ResNet50}. This is because \emph{STC-like} is designed to only allow a maximum of 2$\times$ speedup with the 2:4 sparsity pattern. Furthermore, even if \emph{ResNet50} has $\sim$60\% sparse activations, \emph{STC-like} cannot exploit activation sparsity for speedup. On the other hand, \emph{DSTC-like} is able to translate the sparsity in both weights and activations into reductions in both processing speed and energy consumption. Thus, even if \emph{STC-like} has a low sparsity tax, when running \emph{ResNet50}, its inability to translate various sparsity degrees into hardware savings results in higher EDP than \emph{DSTC-like}.

\textbf{Inefficiency of Unstructured Sparse Designs:} \emph{DSTC-like} is outperformed by \emph{STC-like} on \emph{Transformer-Big}, as shown in Fig.~\ref{fig:hss_mov}. This is because \emph{DSTC-like}'s outer-product-based dataflow with expensive accumulation buffer has a high sparsity tax.  Since \emph{Transformer-Big} has less than 10\% average sparsity in activations, \emph{DSTC-like}'s savings are overshadowed by its hardware support with high sparsity tax. Thus, even though \emph{DSTC-like} has high flexibility, when running \emph{Transformer-Big}, its inefficient sparsity support with high overhead results in higher EDP than \emph{STC-like} does.

\noindent \textbf{Takeaway: there is no existing sparse accelerator that always has lower EDP for both evaluated DNNs because of their respective limitations.}

\subsection{Need for a Flexible and Efficient Design}
\label{sec:existing-hw-support}

To develop a flexible accelerator that is efficient for various DNNs with diverse sparsity degrees, \emph{the community can benefit from a general design that is simultaneously flexible and efficient}. However, as already demonstrated by the \emph{DSTC-like} and \emph{STC-like} comparisons above, many existing sparse DNN designs tend to trade flexibility for efficiency or vice versa, thus facing the challenge of not being able to meet both requirements.

To address this problem, we introduce a hardware-software co-design approach motivated by a novel class of sparsity patterns: \textbf{hierarchical structured sparsity (HSS)}, which leverages multiple levels of G:H structured sparsity to express diverse sparsity degrees in a multiplicative fashion. As shown in Fig.~\ref{fig:hss_mov},  while maintaining accuracy with our HSS-based sparsification, our low-sparsity-tax HSS-based hardware accelerator, \designName,  provides lower EDP in both scenarios.


\section{Precise Sparsity Specification}
\label{sec:sparsity_classification}

In order to clearly compare existing sparsity patterns and distinguish our proposed HSS from existing works, it is necessary to precisely describe various sparsity patterns. However, conventional sparsity pattern classification approaches are often based on names that provide an informal characterization of just the dimensions on which the pattern is imposed, and thus fail to distinguish between different sparsity pattern proposals~\cite{pruning-state-mlsys, sparse-inference-and-training-survey} (\eg,  the term \textit{sub-channel} is repeatedly used to describe many different patterns presented in Table~\ref{tab:sparsity-tree}).

To solve the problem,  we propose a precise way of specifying various sparsity patterns based on the fibertree abstraction~\cite{dnn_syn_lec}. As shown in Table~\ref{tab:sparsity-tree}, our specification can easily distinguish between existing sparsity patterns and cleanly reflects the properties of an example sparsity pattern that belongs to our proposed hierarchical structured sparsity (HSS).

\renewcommand{\arraystretch}{1.2}
\begin{table}[tb]
\centering
\resizebox{\columnwidth}{!}{
\begin{tabular}{c|c|c}
\textbf{\begin{tabular}[c]{@{}c@{}}Example\\ Pattern\end{tabular}} &
  \textbf{\begin{tabular}[c]{@{}c@{}}Conventional \\ Classification\end{tabular}} &
  \textbf{\begin{tabular}[c]{@{}c@{}}Fibertree-based Specification\\ \textit{Rank} ($<$rule$>$)\ldots\end{tabular}} \\ \hline
\cite{deep-compression} &
  Unstructured &
  \multicolumn{1}{c}{\begin{tabular}[c]{@{}c@{}} CRS(Unconstrained)\end{tabular}} \\ \hline
\cite{channel-pruning} (Fig.~\ref{fig:sparse-tree-examples}(a)) &
  \begin{tabular}[c]{@{}c@{}}Channel\end{tabular} &
  \multicolumn{1}{c}{\begin{tabular}[c]{@{}c@{}}C(Unconstrained)$\rightarrow$R$\rightarrow$S\end{tabular}} \\ \hline
\cite{patdnn} &
  \begin{tabular}[c]{@{}c@{}}Sub-kernel\end{tabular} &
  \multicolumn{1}{c}{\begin{tabular}[c]{@{}c@{}}C$\rightarrow$RS(G:H), \emph{with any G, H}\end{tabular}} \\ \hline
\cite{NM-finetune} (Fig.~\ref{fig:sparse-tree-examples}(b)) &
  \begin{tabular}[c]{@{}c@{}}Sub-channel\end{tabular} &
  \multicolumn{1}{c}{\begin{tabular}[c]{@{}c@{}} RS$\rightarrow$C$_1$$\rightarrow$C$_0$(2:4)\end{tabular}} \\ \hline
\cite{vector-sparse-tensor-core} &
  \begin{tabular}[c]{@{}c@{}}Sub-channel\end{tabular} &
  \multicolumn{1}{c}{\begin{tabular}[c]{@{}c@{}}RS$\rightarrow$C$_1$$\rightarrow$C$_0$(4:16)\end{tabular}} \\ \hline
\cite{s2ta} &
  \begin{tabular}[c]{@{}c@{}}Sub-channel\end{tabular} &
  \multicolumn{1}{c}{\begin{tabular}[c]{@{}c@{}}RS$\rightarrow$C$\textsubscript{1}$$\rightarrow$C$\textsubscript{0}$({G$\le$8}:8) \end{tabular}} \\ \hline
\textbf{\begin{tabular}[c]{@{}c@{}}Example \\ Two-rank HSS (Fig.~\ref{fig:2rank-hss-tree})\end{tabular}} &
  \begin{tabular}[c]{@{}c@{}}Sub-channel\end{tabular} &
  \multicolumn{1}{c}{\begin{tabular}[c]{@{}c@{}} RS$\rightarrow$C$\textsubscript{N-1}$$\rightarrow$ \\ C$\textsubscript{N-2}$(3:4)$\rightarrow$...$\rightarrow$C$\textsubscript{0}$(2:4) \end{tabular}} \\
\end{tabular}
}

\caption{\textbf{Informal conventional classification and the precise fibertree-based specifications for example sparsity patterns. For fibertree-based specificatons, ranks without pruning rules, \ie, N/A rules in the figures, do not have (). Partitioned ranks are indicated by appending a number to the rank name, e.g., C is split into C$\textsubscript{1}$ and C$\textsubscript{0}$. Note that there could be multiple G and H values allowed for each rank. } }

\label{tab:sparsity-tree}
 \vspace{-5pt}
\end{table}
\renewcommand{\arraystretch}{1}
\subsection{Fibertree Abstraction}
\label{sec:fibertree_abs}

 The fibertree abstraction~\cite{dnn_syn_lec, drt, teaal} provides a systematic and precise way to express the content of tensors, \emph{without getting into the complexities related to its layout when the tensor is actually stored in buffers (\eg, compressed or uncompressed}). Since the sparsity specification focuses on understanding the nature of the sparsity patterns rather than how they can be compressed, we use fibertrees as a basis for our proposed methodology.  

\begin{figure}[t]%
    \centering
    \includegraphics[width=0.8\columnwidth]{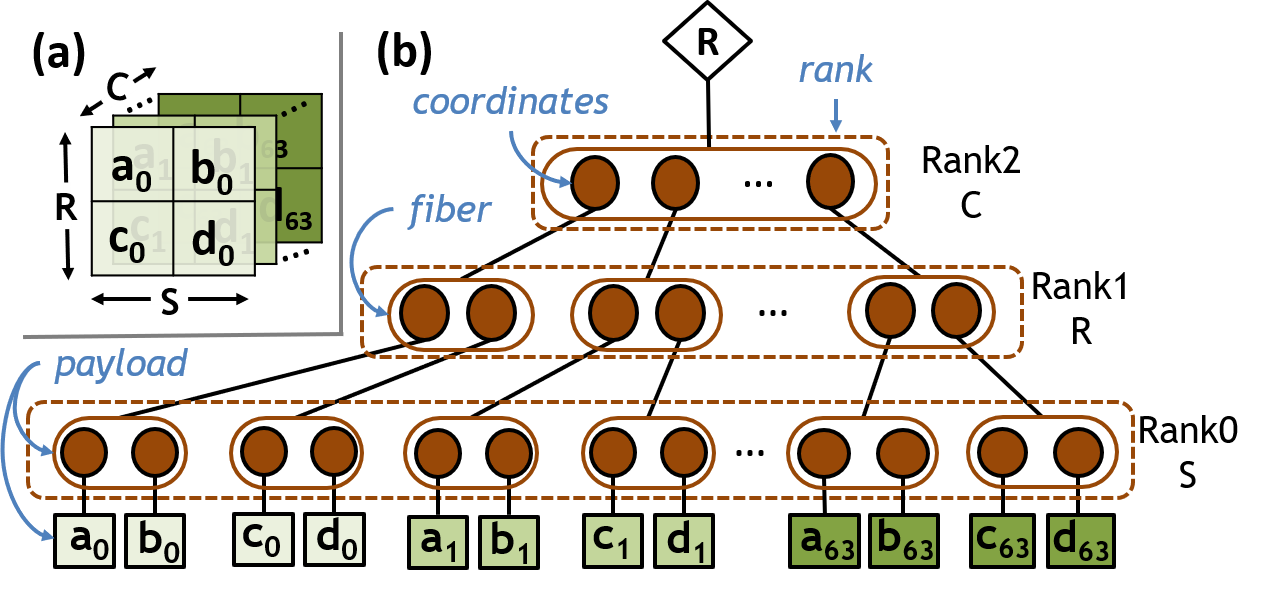} 
    \caption{\textbf{(a) Example dense weight tensor. \textit{C}: channels, \textit{R}: height, \textit{S}: width.  (b) Corresponding fibertree-based abstraction of the tensor. Each tensor dimension corresponds to a level of the tree, referred to as a \textit{Rank}. }}
    \label{fig:tree-repr}
    \vspace{-5pt}
\end{figure}
 
 For ease of presentation, we use the three-dimensional weight tensor in Fig.~\ref{fig:tree-repr}(a) as an example, which has \emph{C} channels, \emph{R} rows, and \emph{S} columns.
Fig.~\ref{fig:tree-repr}(b) shows the fibertree representation of the tensor. The fibertree has three levels, each of which is referred to as a \emph{rank} and corresponds to a dimension of the tensor (\eg, the lowest rank, \emph{Rank0}, corresponds to dimension \emph{S}). Each rank contains multiple \emph{fibers}, each of which contains a set of \emph{coordinates} and their associated \emph{payloads}. For intermediate ranks, the payload is a fiber from a lower rank (\eg, the first coordinate in \emph{Rank1} has the first fiber in \emph{Rank0} as its payload); for a coordinate in \emph{Rank0}, the payload is a simple value (\eg, the first coordinate in \emph{Rank0} has the value $a_{0}$ as its payload. 

\subsection{Fibertree-based Sparsity Specification}
\label{sec:sparsity_specification}



With the fibertree fundamentals presented above, we now discuss how to use such an abstraction to describe sparsity in a tensor. Sparsity is introduced via pruning away the \emph{coordinates} in the dense fibertree. At a high level, to define a specific sparsity pattern, a rank order needs to be specified and each \emph{rank} is assigned a \emph{pruning rule}. The rule specifies if the coordinates in each of its \emph{fibers} can be pruned away; and if so, whether there is a pattern that the per-fiber pruning should follow.

Coordinates in arbitrary ranks can be pruned away. Pruning a coordinate at the lowest rank simply removes values, whereas pruning a coordinate at intermediate ranks removes its fiber payload (\ie, the entire subtree associated with the coordinate), implicitly pruning away all the associated lower-level coordinates. Because of this chained effect, the introduced sparsity is conventionally known as \emph{structured sparsity}. Structured sparsity can have structures at different granularities, which are impacted by the rank at which the pruning rules are defined. 
\begin{figure}[t]%
    \centering
    \includegraphics[width=\columnwidth]{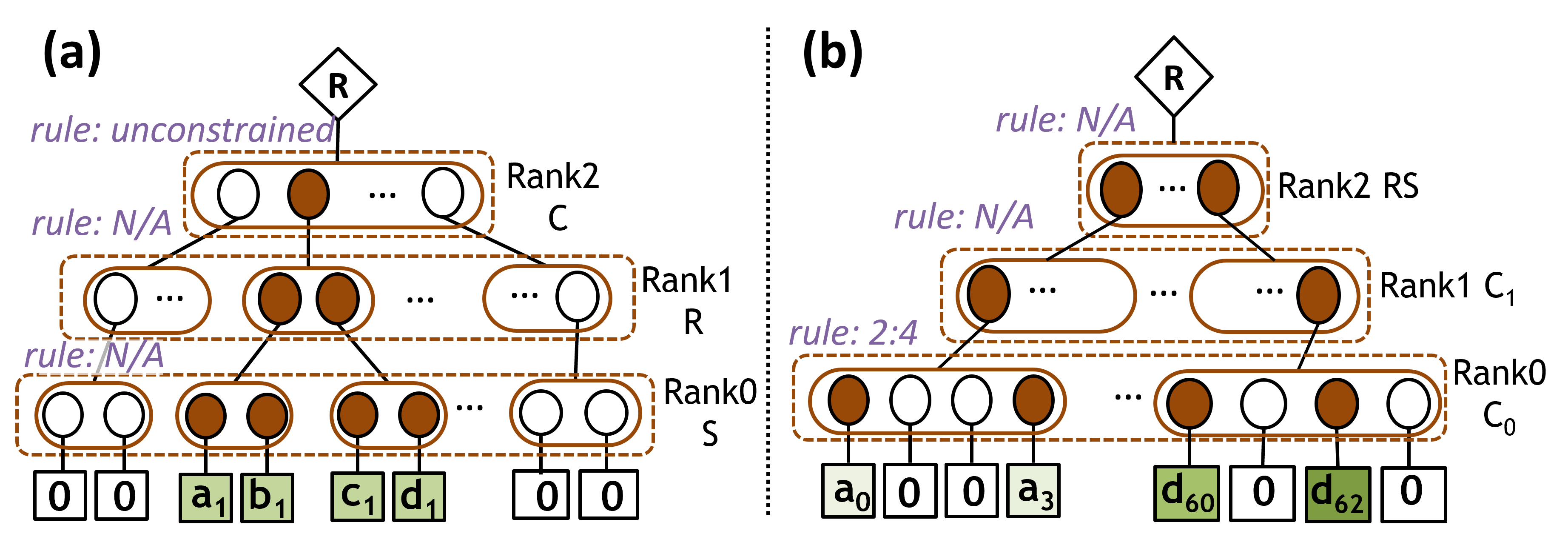} 
    \caption{\textbf{Fibertree-based specification for popular sparsity patterns applied to the tensor in Fig.~\ref{fig:tree-repr}(a). \textbf{(a)} channel-based structured~\cite{channelprune}: C(unconstrained)$\rightarrow$R$\rightarrow$S. \textbf{(b)} 2:4 structured~\cite{NM-finetune}: RS$\rightarrow$C$\textsubscript{1}$$\rightarrow$C$\textsubscript{0}$(2:4). Please note that the abstraction describes the exact sparsity, and is orthogonal to how the tensor is stored in the buffer. }}
    \label{fig:sparse-tree-examples}
    \vspace{-5pt}
\end{figure}
For example, Fig.~\ref{fig:sparse-tree-examples}(a) shows the fibertree-based specification of the conventionally known \emph{channel-based structured sparsity}, which demands arbitrary channels to be completely removed. 
The fibertree-based representation specifies the \emph{unconstrained} pruning rule at the top rank \emph{C}, as each removed coordinate corresponds to a removed channel, which is indicated by the empty circles. We specify such a structure as C(unconstrained)$\rightarrow$R$\rightarrow$S as shown in Table~\ref{tab:sparsity-tree} and Fig.~\ref{fig:sparse-tree-examples}(a). The $\rightarrow$ defines the higher to lower rank order, and ranks with pruning rules carry ($<$rule$>$). The removal of the highest-rank coordinates results in all zeros in the corresponding channels. Since the removal of lower ranks is always implicit, 
the R and S lower ranks are not associated with any explicit pruning rules and are thus not followed by ($<$rule$>$) in the specification.

Furthermore, a sparsity pattern specification may involve first applying \emph{content-preserving transformations} to the tensor, such as \emph{reordering}, \emph{flattening}, or \emph{partitioning} the ranks~\cite{teaal}. For example, this is done for creating the conventionally known sub-channel-based 2:4 structured sparsity~\cite{ampere-white-paper}. Fig.~\ref{fig:sparse-tree-examples}(b) shows the 2:4 structured sparsity that's supported by NVIDIA's sparse tensor core. The original ranks are first reordered to have C as the lowest rank, 
and the C rank is then partitioned into two ranks, C$\textsubscript{1}$ and C$\textsubscript{0}$.
The G:H-style sparsity structure (as described in Section~\ref{sec:sparsity-schemes}) is manifest by allowing at most 2 non-zero values in each fiber of the C$\textsubscript{0}$ rank, which due to the partitioning have exactly four coordinates. For a specific fiber, we refer to the total number of coordinates as its \emph{shape} and the number of coordinates associated with nonzeros as its \emph{occupancy}. Thus, the fiber shape in C$\textsubscript{0}$ rank is defined by the denominator of the fraction, \ie, H in C$\textsubscript{0}$(G:H) structured sparsity, and the max fiber occupancy is defined by the numerator in the fraction. This sparsity pattern is thus specified as RS$\rightarrow$C$\textsubscript{1}$$\rightarrow$C$\textsubscript{0}$(2:4).

As illustrated in Table~\ref{tab:sparsity-tree}, our proposed fibertree-based specification allows precise descriptions of many existing works that were not easily distinguishable using conventional sparsity pattern classification techniques.

\section{HSS}
\label{sec:hss-def}

With the fibertree-based sparsity specifications, it is clear that existing works in Table~\ref{tab:sparsity-tree} all propose to apply different sparsity pattern(s) to \emph{one} rank. A natural approach to enhance such works to represent more sparsity degrees would be to introduce sparsity patterns to \emph{multiple} ranks, motivating our proposed concept of \emph{hierarchical structured sparsity} (HSS).

\subsection{General Concept}

\begin{figure}[t]%
    \centering
    \includegraphics[width=\columnwidth]{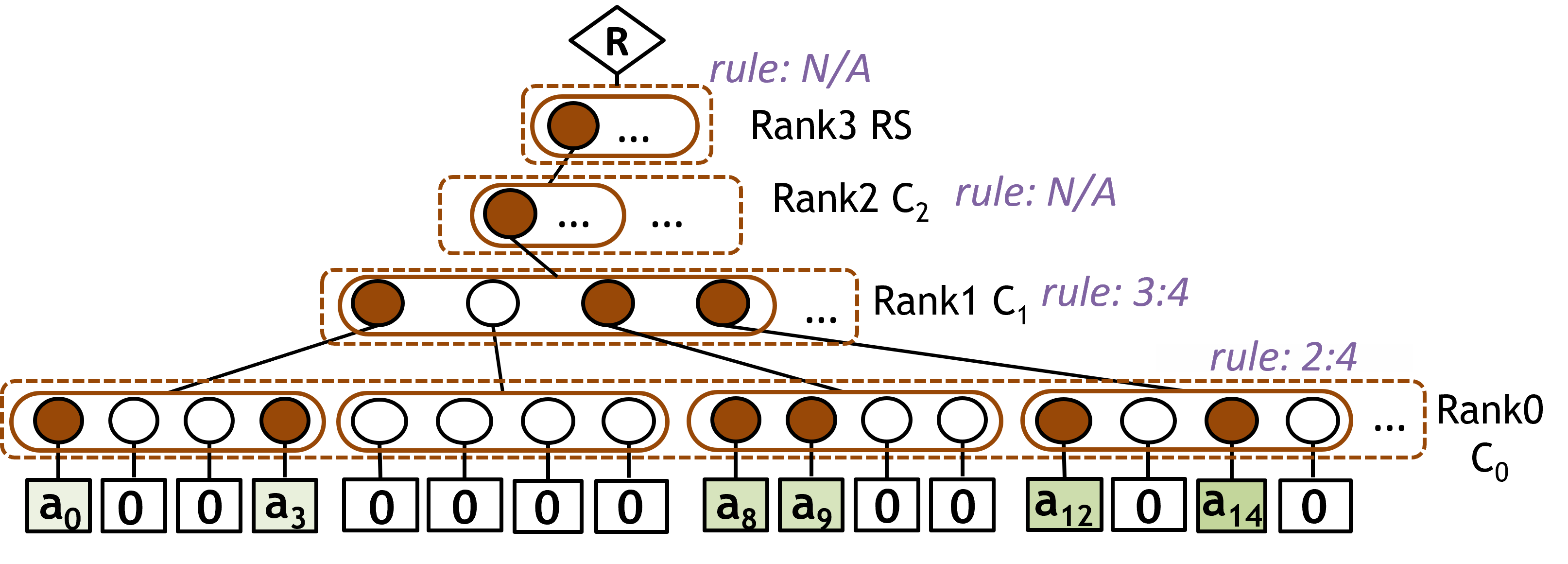} 
    \caption{\textbf{Fibertree-based specification for an example two-rank HSS with RS$\rightarrow$C$\textsubscript{2}$$\rightarrow$C$\textsubscript{1}$(3:4)$\rightarrow$C$\textsubscript{0}$(2:4). Please note that, for general HSS patterns, the choices of number of ranks, rank ordering, flattening, and splitting are not limited to the ones presented in this example.}}
    \label{fig:2rank-hss-tree}
    \vspace{-5pt}
\end{figure}

HSS allows \emph{multiple ranks} to be sparse, where each rank has its own sparsity pattern(s). Such a hierarchy of sparsity patterns could lead to structured sparsity with more sparsity degrees than only one rank with a very flexible sparsity pattern. Since HSS allows more than one rank to be assigned with sparsity patterns, we introduce the parameter N which describes \emph{the number of ranks with sparsity patterns assigned}. For each rank n where 0$\le$n$\le$N-1 and N$\geq$1, a G:H pattern is assigned. G:H ratios can be different for different ranks. We call an instance of HSS that contains N ranks with sparsity patterns assigned an N-rank HSS (\eg, in Fig.~\ref{fig:2rank-hss-tree}, there are four ranks in the fibertree, and two of them have dedicated G:H sparsity patterns, so we call it a two-rank HSS).

\subsubsection{Fibertree-based Specification}
To more concretely illustrate the idea of HSS, we will refer to the fibertree-based specification of the two-rank HSS pattern shown in Fig.~\ref{fig:2rank-hss-tree}. However, for general HSS patterns, the choices of rank ordering, flattening, and splitting are not limited to the ones shown in this example. 

The example HSS pattern orders the ranks in R, S, C fashion, similar to the one in Fig.~\ref{fig:sparse-tree-examples}(b). Unlike in Fig.~\ref{fig:sparse-tree-examples}(b), where the original C rank is partitioned into two ranks, the example HSS pattern partitions the original C rank into three ranks C$\textsubscript{2}$, C$\textsubscript{1}$, and C$\textsubscript{0}$, and assigns 3:4 and 2:4 to the lowest two ranks, \ie,  C$\textsubscript{1}$ and C$\textsubscript{0}$. Such a two-rank HSS pattern can be described as RS$\rightarrow$C$\textsubscript{2}$$\rightarrow$C$\textsubscript{1}$(3:4)$\rightarrow$C$\textsubscript{0}$(2:4). As shown in Table~\ref{fig:sparse-tree-examples}, the specification has more than one rank with pruning rules assigned, resulting in qualitatively different sparsity patterns compared to existing DNN sparsity patterns. 

\begin{figure*}[tb]%
    \centering
     \begin{subfigure}[b]{0.97\columnwidth}
         \includegraphics[width=\columnwidth]{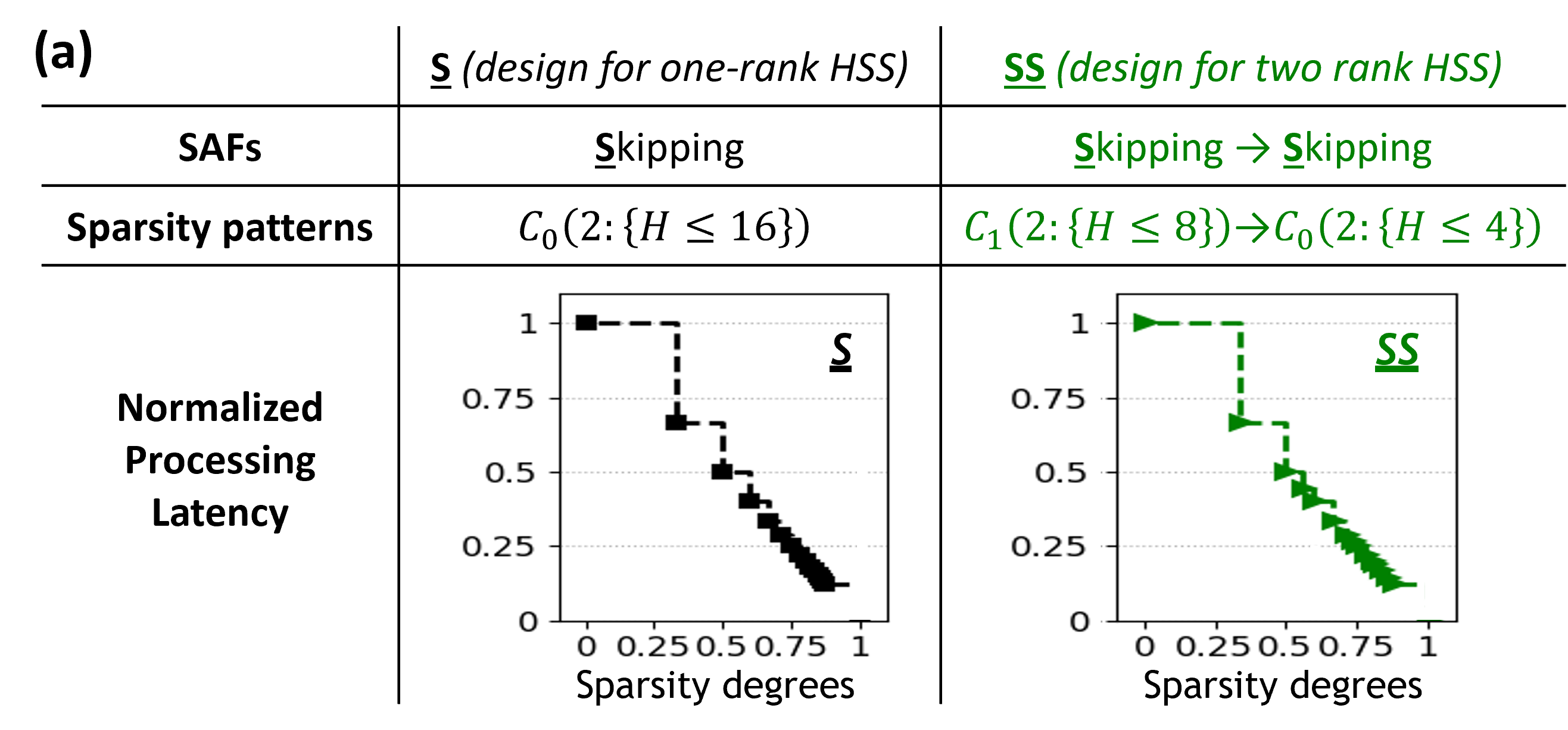}
     \end{subfigure}
     \hfill
    \begin{subfigure}[b]{0.28\columnwidth}
         \includegraphics[width=\columnwidth]{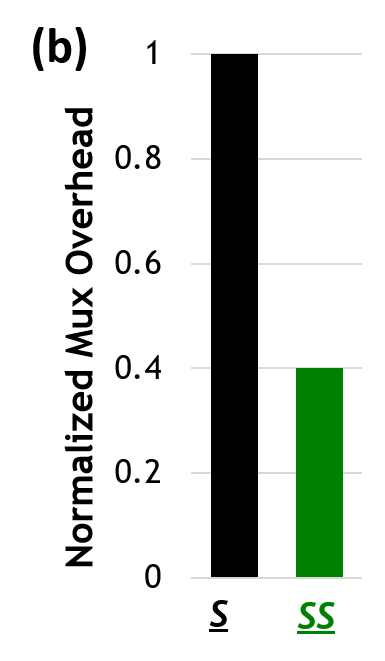}
     \end{subfigure}
    \hfill
     \begin{subfigure}[b]{0.75\columnwidth}
         \includegraphics[width=\columnwidth]{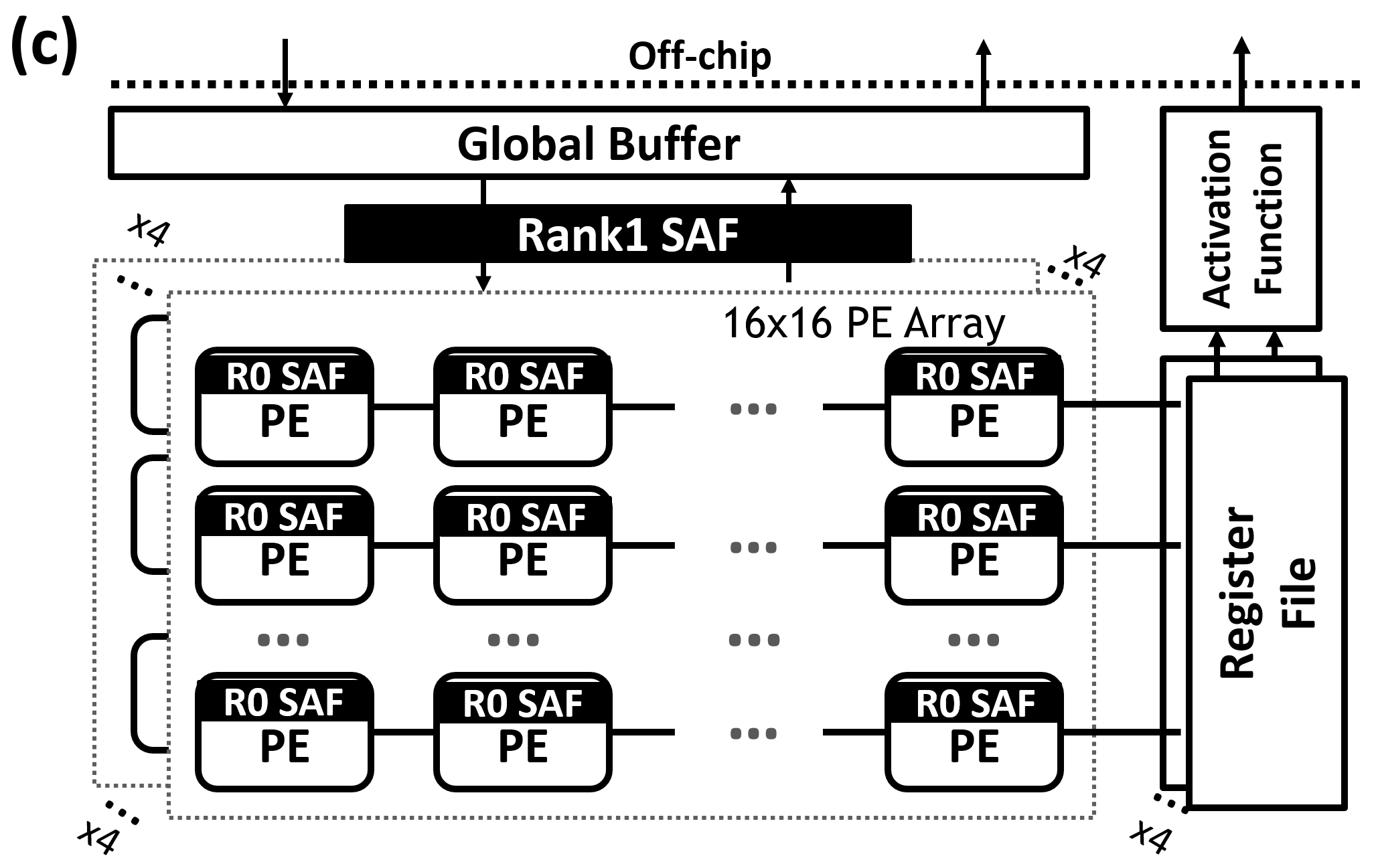}
     \end{subfigure}
\caption{\textbf{Comparison of designs with the same flexibility (15 sparsity degrees across 0\%-87.5\%) but different numbers of ranks. \textcolor{Green}{\underline{SS}} shows great potential for high flexibility and efficiency. (a) Design attributes and normalized processing latency (markers indicate the discrete sparsity degrees.) (b) Normalized muxing overhead. (c) High-level architecture of \designName, with modularized SAFs for each rank at different architecture levels.}}
\label{fig:hss-repr-compare}
\end{figure*}


\subsubsection{Sparsity Degrees}

Different ranks in a multi-rank HSS have sparsity patterns with different granularity. For example, in Fig.~\ref{fig:2rank-hss-tree}, the lowest C$\textsubscript{0}$ rank's 2:4 structure is based on a single-value granularity. Whereas the higher C$\textsubscript{1}$ rank's 3:4 structure is based on a larger granularity that's the shape of the lower fiber. Thus, the 3:4 ratio describes whether a fiber payload of each coordinate must contain all zeros or can contain nonzeros.

The overall sparsity degree of an HSS tensor can be derived from its sparsity structures at each rank.  For example, the two-rank HSS RS$\rightarrow$C$\textsubscript{2}$$\rightarrow$C$\textsubscript{1}$(3:4)$\rightarrow$C$\textsubscript{0}$(2:4) in Fig.~\ref{fig:2rank-hss-tree} has a sparsity of  $1-\frac{3}{4}\times\frac{2}{4}=0.625$. In general, the overall sparsity degree can be expressed as $sparsity =  1- \prod_{n=0}^{N-1}{\frac{G_n}{H_n}}$, where G$\textsubscript{n}$:H$\textsubscript{n}$ is the ratio assigned to rank $n$. Thus, by assigning a different number of ranks and different G:H ratios at each rank, HSS allows a flexible and systematic expression of various overall sparsity degrees.

\noindent\textbf{Note: for ease of presentation, we will succinctly specify all sparsity patterns with only the ranks with sparsity patterns, \eg, RS$\rightarrow$C\textsubscript{1}$\rightarrow$C$\textsubscript{0}$(2:4) is simplified to C$\textsubscript{0}$(2:4)}.

\subsection{DNN Sparsification with HSS}
\label{sec:hss_sparsification}

Similar to all existing DNN sparsity patterns, HSS patterns can be introduced into DNNs to produce sparse DNN models. The goal for HSS-based sparsification is to ensure the most important nonzero values are preserved as much as possible.

To achieve the goal, we sparsify a dense tensor rank-by-rank in a lower-to-higher fashion. For example, for a C$\textsubscript{1}$(3:4)$\rightarrow$C$\textsubscript{0}$(2:4) HSS, we first apply the rank C$\textsubscript{0}$'s 2:4 pattern and then rank C$\textsubscript{1}$'s 3:4 pattern. For the lowest rank, we sparsify the values with the smallest magnitude. For an intermediate rank, we prune coordinates whose fiber payload has the smallest \emph{scaled L2 norm}, defined as the average magnitude of all values in the payload. Depending on the per-rank sparsity patterns, the flexibility of HSS allows us to obtain sparse models with diverse sparsity degrees.

Since the introduced sparsity pattern is orthogonal to the pruning algorithm choice (\eg, pruning on trained dense model~\cite{NM-finetune}, pruning from scratch~\cite{NM-training}, pruning with value revival~\cite{prune-transfomer-revive}, etc.), it is the algorithm designer's freedom to decide whether the ranks are sparsified at once or gradually sparsified over the process. As we will show in Sec.~\ref{sec:results}, even with a traditional pruning algorithm, a DNN with HSS patterns can maintain reasonable accuracy.


\section{\designName \ Overview}
\label{sec:higlight_overview}

HSS unveils an organized HSS-based hardware design space, where each design can be systematically developed by considering three aspects for each HSS operand:
\begin{itemize}[topsep=0pt,itemsep=0pt,partopsep=0pt,parsep=0pt, leftmargin=12pt]
    \item G:H patterns supported at a rank. 
    \item the number of HSS ranks supported by the hardware.
    \item the acceleration techniques, referred to as \emph{sparse acceleration features (SAFs)}~\cite{micro-2022-sparseloop}, supported at each rank.
\end{itemize}
In this section, we motivate \designName's high-level architecture by discussing the impact of making different design decisions for the above design aspects.


\subsection{Impact of Supported SAF at Each Rank}
The accelerator can have different supported SAFs at a rank to translate sparsity into different savings. Specifically, when there are ineffectual operations, the hardware can employ
\begin{itemize}[topsep=0pt,itemsep=0pt,partopsep=0pt,parsep=0pt, leftmargin=12pt]
    \item \textbf{Gating:} lets the hardware stay idle to save energy. Gating often involves a trivial sparsity tax, \eg, an AND gate.
    \item \textbf{Skipping}: fast forwards to the next effectual operation to save energy and time. Skipping incurs a higher sparsity tax, \eg, muxing logic for leader-follower intersections.
\end{itemize}
Since gating is undesirable for many latency-sensitive applications, we will focus on discussing the impact of various design aspects assuming skipping as the supported SAF. 

Skipping is highly reliant on high utilization of the components to achieve the desired speedup, so it is desirable to support G:H patterns with a fixed (set of) G that is a factor of the number of parallel hardware units (\eg, with four processing elements (PEs), it is desirable to support G=4 patterns, as all PEs can be utilized with the four nonzeros in the block of H values, regardless of what H is). As a result, as shown in Fig.~\ref{fig:hss-repr-compare}(a), the example designs with skipping SAFs support sparsity patterns with a fixed G value of two at each rank.

\noindent\textbf{Takeaway: it's more desirable to support skipping, which favors G:H patterns with a G that's a factor of the available number of hardware instances to more easily ensure workload balancing with a high hardware utilization.}

\begin{figure}[tb]%
   \centering
   \includegraphics[width=\columnwidth]{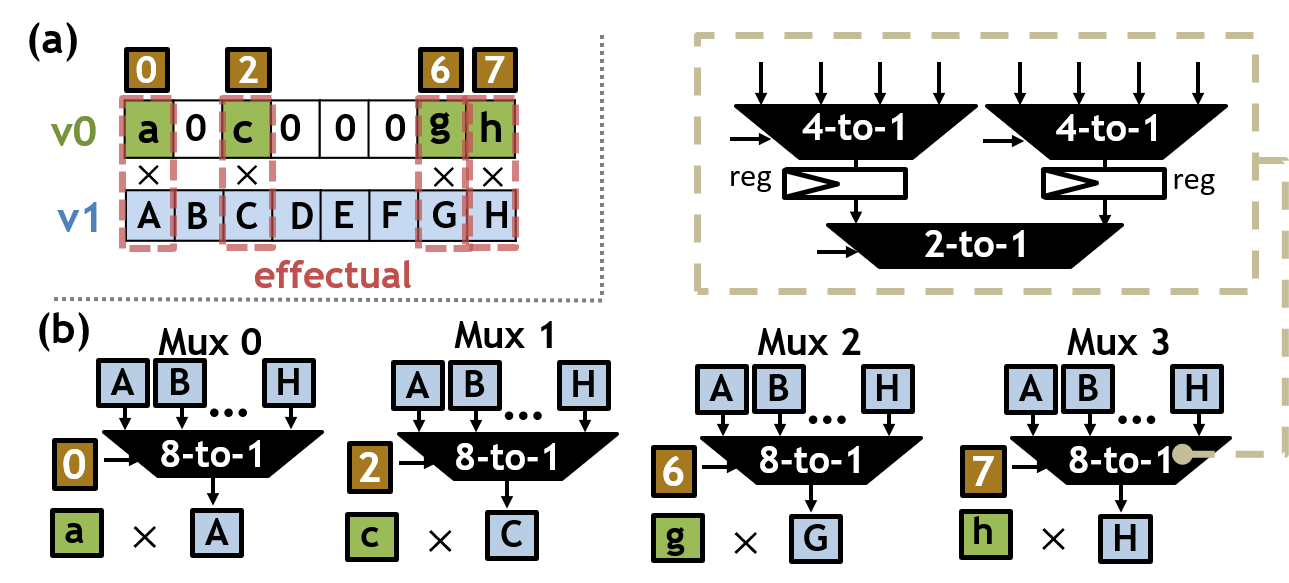}
   \caption{\textbf{ (a) Effectual computations in a dot product with 2:4 sparse vector 0 (v0) and dense vector 1 (v1). (b) Implementation of the muxing logic for selecting four values from a block of eight values.}}
   \label{fig:muxes}
   \vspace{-5pt}
\end{figure}

\subsection{Impact of Per-rank Supported Patterns}
 
To implement skipping for a G:H pattern, additional hardware is needed. For example,  Fig.~\ref{fig:muxes}(a) shows a dot product workload with two vectors: vector 0 (\emph{v0}) is 4:8 sparse, and vector 1 (\emph{v1}) is dense.  In order to \emph{only} perform effectual computations, \ie, skip the ineffectual computations, the accelerator needs 8-to-4 muxing logic to select the correct \emph{v1} values. Specifically, as shown in Fig.~\ref{fig:muxes}(b), the 8-to-4 muxing logic can be implemented with four 8-to-1 muxes. For example, the first 8-to-1 mux selects {\small\colorbox{cyan}{{A}}} based on {\small\colorbox{SpringGreen}{{a}}}'s coordinate {\small\colorbox{brown}{\textcolor{white}{0}}}. To ensure low latency, an 8-to-1 mux can be implemented with two 4-to-1 muxes pipelined with a 2-to-1 mux. 
 
 Since an accelerator can be designed to support multiple G:H patterns with different H values, the muxing sparsity tax increases as the largest supported H value, \ie, H\textsubscript{max}, increases. Specifically, in order to support all possible patterns, the accelerator needs G number of H\textsubscript{max}-to-1 muxes. 

\noindent\textbf{Takeaway: with a fixed G, the energy and area sparsity tax increases approximately linearly with H\textsubscript{max}.}

\subsection{Impact of Supported Number of Ranks}

Given a target number of sparsity degrees to represent, supporting more HSS ranks reduces the H\textsubscript{max} at each rank, reducing the sparsity tax. Specifically, due to the nature of fraction multiplications, multi-rank HSS can easily represent a large number of sparsity degrees with a much smaller H\textsubscript{max} at each rank by exploiting the composability of sparsity patterns. As shown in Fig.~\ref{fig:hss-repr-compare}(a), with both designs supporting 15 different sparsity degrees across 0\% to 87.5\%, compared to the one-rank HSS design \underline{S}, which requires a H\textsubscript{max} of 16, the two-rank HSS design \textcolor{Green}{\underline{SS}} only requires H\textsubscript{max} of 8 at Rank1 and a H\textsubscript{max} of 4 at Rank0. 

The sparsity support of a multi-rank HSS design is implemented at different architecture levels, each of which targets a specific rank (\eg, in Fig.~\ref{fig:hss-repr-compare}(c), the processing element (PE) array level implements \textit{Rank1 SAF} to exploit Rank1 patterns and the PE level implement \emph{Rank0 SAF} to exploit Rank0 patterns). Fig.~\ref{fig:hss-repr-compare}(b) shows the normalized sparsity tax for the two HSS designs in terms of their muxing overhead. Due to the reduced H\textsubscript{max} values at each rank, \textcolor{Green}{\underline{SS}} introduces $>2\times$ less muxing overhead, \ie, lower sparsity tax, while representing the same number of sparsity degrees as \underline{S}. 


\noindent\textbf{Takeaway: Compared to the popular one-rank HSS supported in many existing works, hardware designed for multi-rank HSS can represent the same number of sparsity degrees with much lower sparsity tax.}

\subsection{High Level Architecture}
\label{sec:highlight_summary}
 Fig.~\ref{fig:hss-repr-compare}(c) shows the high-level architecture organization of our proposed  \designName \ accelerator, a simultaneously efficient and flexible accelerator consisting of a memory hierarchy and 1024 \emph{MAC}s grouped into four PE arrays.  \designName \ supports DNNs with two-rank HSS weights and unstructured sparse input activations. In terms of SAF choices, \designName \ implements modularized skipping SAFs at different architecture levels to translate the two-rank HSS into energy and latency savings, and performs gating on input activation's sparsity to further reduce energy consumption. 
 

\section{A Deeper Dive Into \designName}
\label{sec:highlight_detail}
In this section, we present more details on \designName's micro-architecture implementations.  \textbf{For ease of presentation, we will use a down-sized architecture with two PEs and sparsity support for C\textsubscript{1}$($2:\{2$\le$H$\le$4\}$)$$\rightarrow$C\textsubscript{0}(2:4) to discuss the core ideas of the \designName \ micro-architecture. }

\subsection{DNNs Processed as Matrix Multiplication}
\label{sec:dnn_as_mm}
\begin{figure}[tb]%
    \centering
    \includegraphics[width=\linewidth]{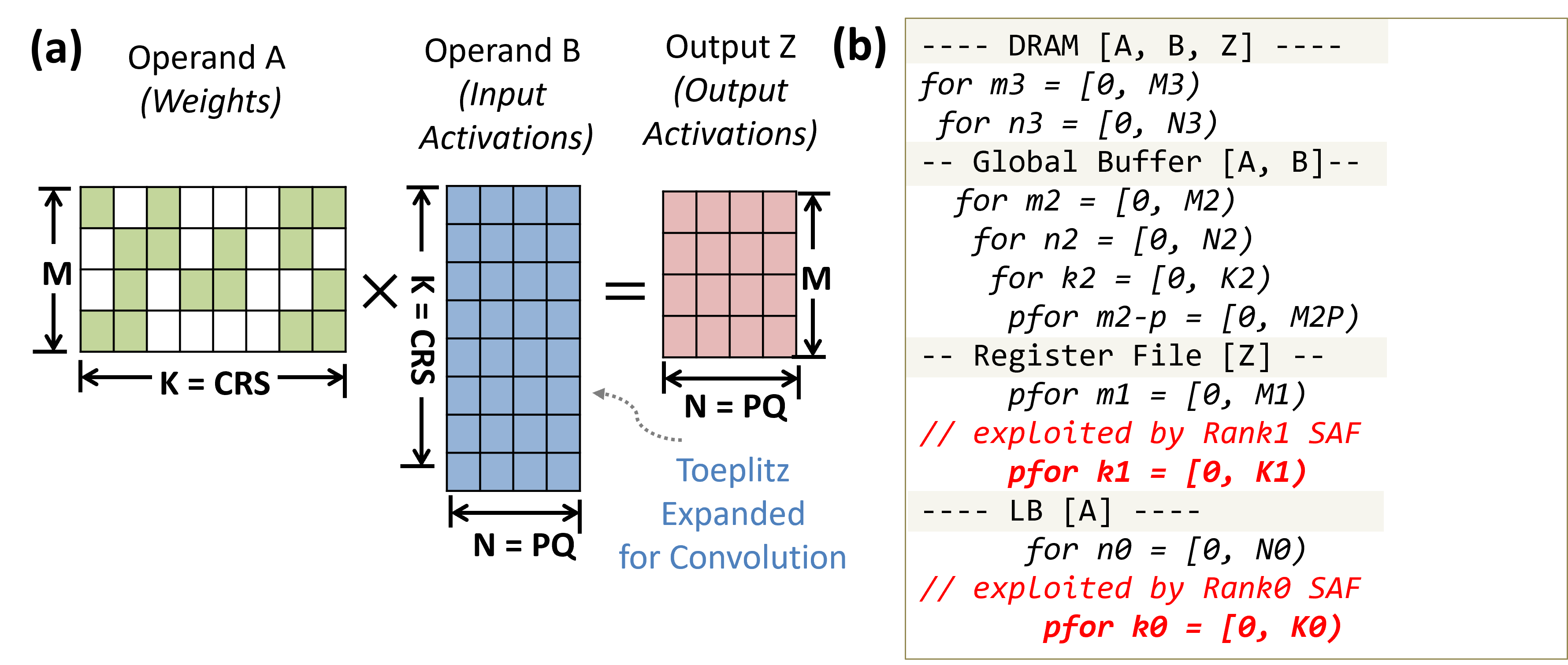} 
    \caption{ \textbf{\textbf{(a)} Convolution represented as matrix multiplication with flattened operand A (weights) and Toeplitz expanded operand B (input activations)~\cite{dnn_syn_lec}. \textit{M: number of filters; C: \# of channels; R,S: height and width of filter kernels; P,Q: height and width of outputs.} \textbf{(b)} Loopnest representation of \designName's dataflow.}}
    \label{fig:toeplitz-mm-loopnest}
\end{figure}

We design \designName \ to process various layers in DNNs as matrix multiplications (MM) workloads (as shown in Fig.~\ref{fig:toeplitz-mm-loopnest}(a), convolutional layers are flattened with Toeplitz expansion on the inputs~\cite{dnn_syn_lec} before sending to the accelerator for processing), as many existing works do~\cite{ampere-white-paper, vector-sparse-tensor-core, s2ta, tpu-datacenter, volta-white-paper, sdp}. Thus, DNN layers with MM kernels (\eg, fully connected layers) are represented as they originally are. Whereas, as shown in Fig.~\ref{fig:toeplitz-mm-loopnest}(a), convolutional layers are represented as MM by flattening the weight dimensions and performing a Toeplitz expansion on the inputs~\cite{dnn_syn_lec} before sending to the accelerator for processing. 
Processing all layers as matrix multiplications implies interchangeable operands. Hence, instead of referring to the operands as weights and input activations, we refer to them as operands A and B, where operand A is dense or HSS, and operand B is either dense or unstructured sparse.

\subsection{Compression Format for HSS} 
\begin{figure}[tb]
    \centering
    \includegraphics[width=0.9\linewidth]{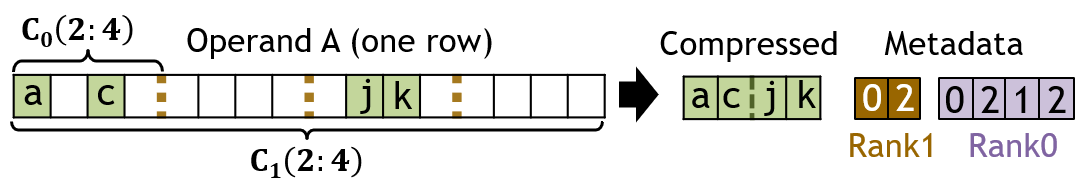} 
    \caption{\textbf{ Hierarchical CP compression for operand A row.} }
    \label{fig:hierarchical-cp}
\end{figure}

To correctly eliminate ineffectual hardware operations, \ie, buffer accesses and computes associated with zeros, it is important to capture both ranks' sparsity structure with metadata. \designName \ uses an offset-based coordinate representation (CP)~\cite{micro-2022-sparseloop} format to describe the position of nonzero values/non-empty blocks at each rank. Fig.~\ref{fig:hierarchical-cp} shows the metadata for an example C$\textsubscript{1}$(2:4)$\rightarrow$C$\textsubscript{0}$(2:4) operand A tensor. For Rank0, each nonzero value carries a CP to indicate its position in its block of  H\textsubscript{0} values (\eg, since {\small\colorbox{SpringGreen}{a}} is at the first position in its block, it carries a  {\small\colorbox{Orchid}{0}} metadata). For Rank1, each nonzero block carries a CP to indicate its relative position in the H\textsubscript{1} blocks (\eg, the first and third blocks have nonzeros and thus carry upper-level metadata {\small\colorbox{brown}{\textcolor{white}{0}}} and {\small\colorbox{brown}{\textcolor{white}{2}}}.) 

\subsection{Hierarchical Skipping}

\begin{figure}[t]%
    \centering
    \includegraphics[width=\linewidth]{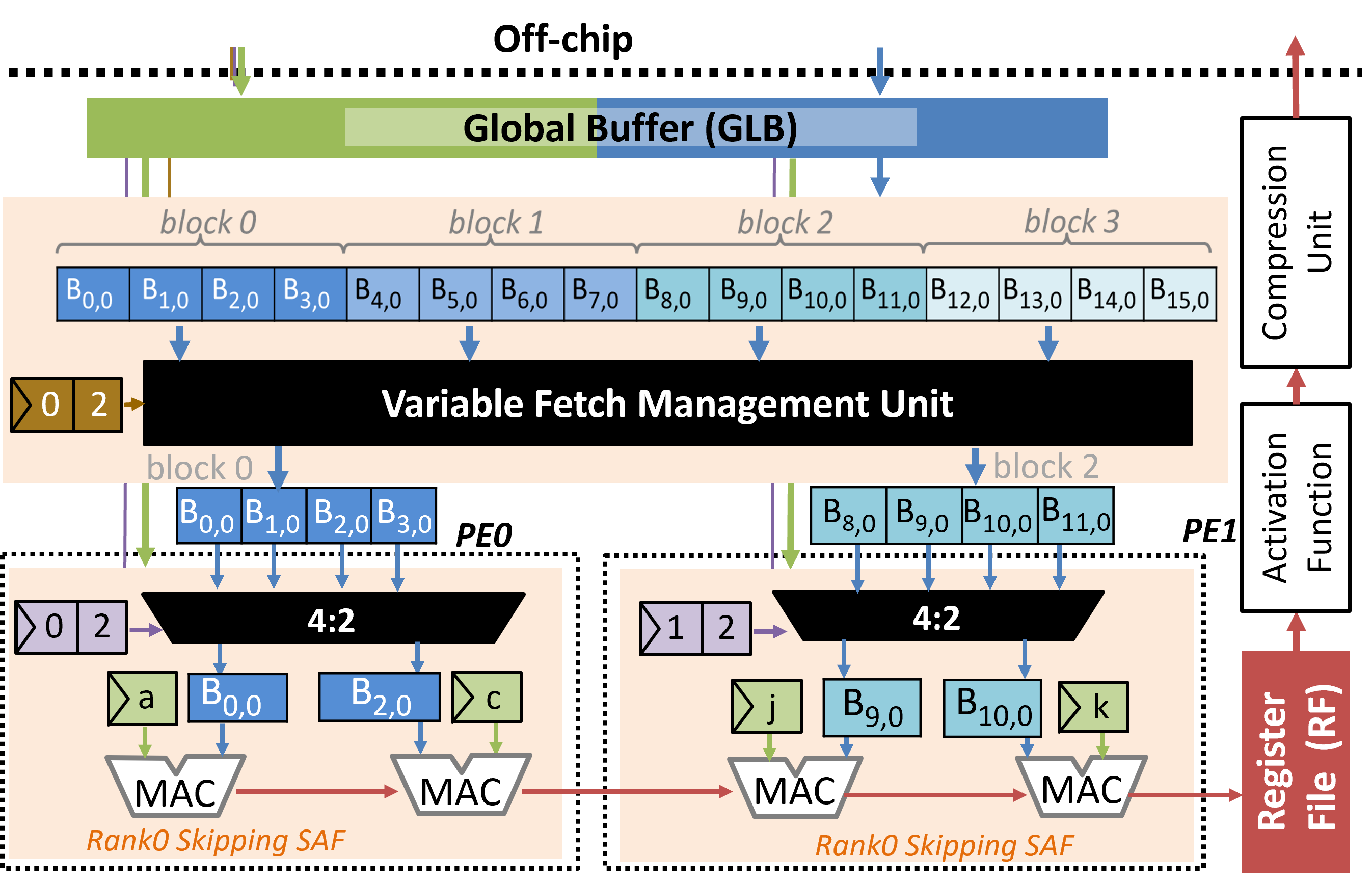} 
    \caption{ Down-sized architecture organization of \designName \ with hierarchical skipping SAF. The showcased processing flow is for the example \emph{C$\textsubscript{1}$(2:4)$\rightarrow$C$\textsubscript{0}$(2:4)} operand A in Fig.~\ref{fig:hierarchical-cp} and a dense operand B. Matched capitalized and lower case letters indicate corresponding values. Boxes with triangles are registers.}
    \label{fig:hierarchical-safs}
\end{figure}

To achieve high utilization of the hardware, and thus fast processing speed, \designName \ employs a hierarchical skipping technique, \ie, both \emph{Rank1 SAF} and \emph{Rank0 SAF} perform skipping based on their target rank's sparsity structure, as shown in Fig.~\ref{fig:hierarchical-safs}. \textbf{Thus, \designName's total speedup is the product of the speedup introduced at each rank.} To illustrate the ideas, we use the C$\textsubscript{1}$(2:4)$\rightarrow$C$\textsubscript{0}$(2:4) operand A shown in Fig.~\ref{fig:hierarchical-cp} and a dense operand B as an example workload. We will discuss sparse B operand support in Sec.~\ref{sec:exploit_sparse_b}.

\subsubsection{HSS-Operand Stationary Dataflow} 
Before diving into the SAFs, we discuss the general processing flow of \designName \ by presenting its \emph{dataflow}, which defines an accelerator's scheduling of data movement and compute in space and time~\cite{eyeriss-dataflow,timeloop-ispass}. 
To exploit the statically known sparsity structure to introduce desirable workload balancing, \designName \ employs an HSS-operand stationary dataflow, where each Rank0 block of A is held stationary in each PE for reuse across different operand B values. As shown in Fig.~\ref{fig:hierarchical-safs}, PE0 holds stationary in registers the two nonzero values \colorbox{SpringGreen}{{a}}, \colorbox{SpringGreen}{{c}} in the first block of operand A. Each MAC in the PE is responsible for working on one of the G nonzeros in its assigned block, specifically, the MAC on the left in PE0 works on \colorbox{SpringGreen}{{a}}, and the right MAC works on \colorbox{SpringGreen}{{c}}. The partial sums calculated at each MAC are first spatially accumulated across the PEs in the same row and updated to the \emph{Register File}.  More dataflow details are described in Fig.~\ref{fig:toeplitz-mm-loopnest}(b) based on the well-known loopnest representation~\cite{timeloop-ispass, eyeriss-dataflow, dnn_syn_lec}.

\subsubsection{Skipping SAF at Rank1} 
\designName's \emph{Rank1 Skipping SAF} exploits the sparsity structure in \emph{Rank1} only. Specifically, it is responsible for only distributing non-empty Rank1 blocks in operand A and the corresponding blocks in operand B to the PEs for parallel processing, \eg,  as shown in Fig.~\ref{fig:hierarchical-safs}, only the nonzeros in the first block (\ie, \colorbox{SpringGreen}{{a}}, \colorbox{SpringGreen}{{c}}) and the third block (\ie, \colorbox{SpringGreen}{{j}}, \colorbox{SpringGreen}{k}) in operand A are transferred to be processed at the PEs. Since only half of the \emph{Rank1} blocks are non-empty in the example tensor, \emph{Rank1 Skipping SAF} introduces a 2$\times$ speedup. Other sparsity patterns (degrees) in rank 1 can be exploited similarly to achieve different speedups. 

Recall that \emph{Rank1 Skipping SAF} is required to support sparsity patterns defined by C$\textsubscript{1}$$($2:\{2$\le$ H$\le$4\}).  To maintain high utilization for different sparsity patterns (degrees), each of the two \emph{PE}s in Fig.~\ref{fig:hierarchical-safs} should always get a non-empty block of operand A.  To ensure correctness, different operand B data need to be selected for computation for different supported sparsity patterns. For different H values at Rank1, a different number of operand B blocks need to be selected from at each processing step. For example, as shown in Fig.~\ref{fig:hierarchical-safs}, four blocks (\emph{b$\textsubscript{0}$, b$\textsubscript{1}$, b$\textsubscript{2}$, and b$\textsubscript{3}$})  need to be selected for operand A having HSS patterns with H$=$4).  


\begin{figure}[tb]%
    \centering
    \includegraphics[width=\linewidth]{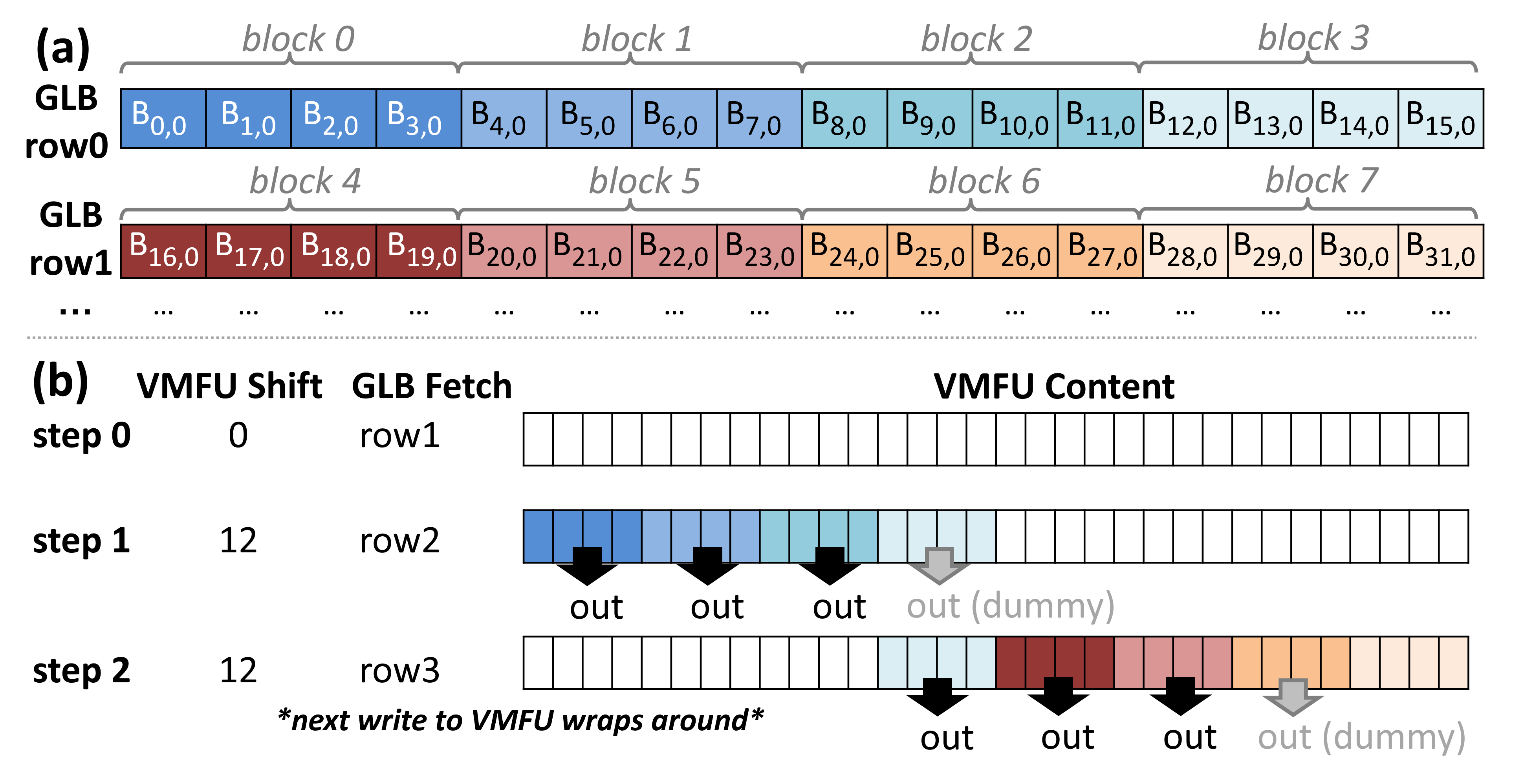}
    \caption{\textbf{(a) Operand B memory layout in GLB. (b) Operand B datamovement between GLB and \emph{Variable Fetch Management Unit (VFMU)}  for the first three processing steps when operand A has a $C_1$\emph{(2:3)} sparsity. To output the correct operand B blocks, \emph{VFMU} is configured to shift by 12 values (three blocks) per read. \textit{out} points to blocks that VFMU outputs to the muxing logic in the step. }}
    \label{fig:denseB-vmfu-23}
\end{figure} 
However, due to the fixed physical dimensions of the GLB, each GLB fetch has to be fixed to a certain number of blocks. As shown  in the example GLB memory layout in Fig.~\ref{fig:denseB-vmfu-23}, each GLB row contains 16 data words from a dense operand B (\ie, four Rank1 blocks). To avoid unaligned \emph{GLB} fetches for different H$\textsubscript{1}$ values, as shown in Fig.~\ref{fig:hierarchical-safs}, \designName's \emph{Rank1 Skipping SAF} employs a \emph{Variable Fetch Management Unit} (\emph{VFMU}) to allow variable length streaming access, which is a technique commonly used for bitstream parsing (\eg, in the entropy coding for video compression~\cite{pipelined-CABAC, parallel-CABAC}). More specifically, \emph{VFMU} includes a small buffer that stores the 2$\times$H$\textsubscript{max}$ blocks of operand B. The buffer is written with data that are fetched from \emph{GLB} in an aligned fashion and can be configured with a \textit{shift} signal to determine the offset position for the current read to start. Fig.~\ref{fig:denseB-vmfu-23} describes the data movement between GLB and \emph{VMPU} for the first three processing steps when operand A has a  G:H$=$2:3 sparsity at C$\textsubscript{1}$. The \emph{shift} signal is configured to three blocks (\ie, 12 values) to allow the correct operand B blocks to be read out. Note that to keep the output width uniform, there are always four blocks read out of the VMPU. However, in the case of G:H$=$2:3, the last block is just a dummy padding that will never be selected by the muxing logic in \emph{Rank1 Skipping SAF}. The \emph{VMPU} processing is trivial to show for operand A with 
G:H$=$2:4, as all accesses are well aligned. Such variable fetch support allows correct operand B to be fetched for different 2:H structures. 


With the correctly shifted blocks, to avoid implementing wide muxes that select from the entire blocks of data, VFMU employs 4-to-2 muxes to select the correct pair of start and end addresses using the metadata from operand A. The addresses are used to index into VFMU's internal registers.
 
\subsubsection{Skipping SAF at Rank0} 
As discussed above, each \emph{PE} in \designName \ works on a non-empty \emph{Rank1} block with \emph{C$\textsubscript{0}$(2:4)}.  To keep the two \emph{MACs} busy in each \emph{PE}, \designName \ employs Rank0 skipping SAF with  4-to-2 muxing logic. As shown in Fig.~\ref{fig:hierarchical-safs}, based on \emph{Rank0}'s CP metadata, the 4:2 mux in each \emph{PE} selects the correct operand B for each \emph{MAC}.

\subsection{Exploiting Operand B Sparsity}
\label{sec:exploit_sparse_b}

So far, we have used a dense operand B in our example workloads to illustrate \designName's processing flow. However, in DNN workloads, operand B for can be sparse due to non-linear activation functions (\eg, ReLU) and/or activation pruning~\cite{s2ta, DASNet-iact-pruning}. \designName \ exploits unstructured sparse operand B through compression and gating.

\begin{figure}[tb]%
    \centering
    \includegraphics[width=\linewidth]{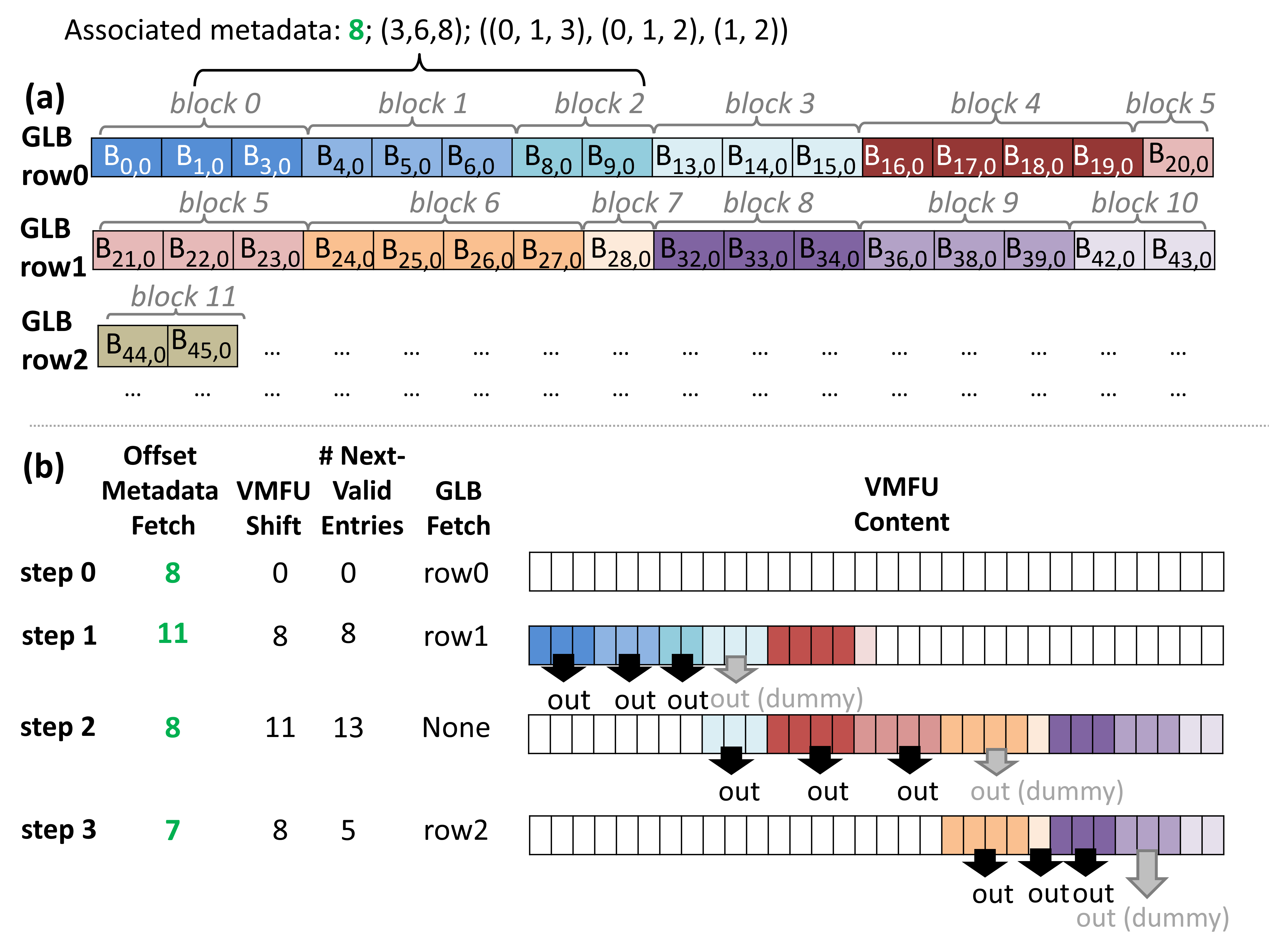}
    \caption{\textbf{(a) Compressed operand B memory layout in GLB and the metadata associated with the first three blocks. (b) Operand B datamovement between GLB and VFMU for the first four processing steps when operand A has a $C-1$\emph{(2:3)} sparsity. VFMU shifts based on the number of nonzeros encoded in the metadata for each set of Rank1 blocks. GLB fetches are only performed when the number of valid data in VFMU cannot meet the need for the next set of Rank1 blocks.} }
    \label{fig:sparseB-vmfu-23}
\end{figure}

We again use the \emph{C$\textsubscript{1}$(2:3)}$\rightarrow$\emph{C$\textsubscript{0}$(2:4)} example to present the hardware support. As shown in Fig.~\ref{fig:sparseB-vmfu-23}(a), when compressed, only the nonzero Operand B values are stored in the GLB. Operand B carries three levels of metadata that hierarchically encodes the nonzero value locations. Specifically, the metadata includes: \textbf{(1)} the total number of nonzeros for every set of Rank1 blocks (three in \emph{C$\textsubscript{1}$(2:3)}); \textbf{(2)} end addresses of each Rank1 block; \textbf{(3)} the intra-Rank0-block offset for each nonzero value. For intermediate layers, such compression on a previous layer's output activation is performed by the compression unit after the activation function unit in Fig.~\ref{fig:hierarchical-safs} to prepare for the processing for the next layer.

 Since different Operand B blocks have different occupancy, instead of always assuming a fixed shift amount as in the case of a dense operand B, the VFMU assigns the shift to be the encoded offset for each set of Rank1 blocks. For example, as shown in Fig.~\ref{fig:sparseB-vmfu-23}(b), the shift at \emph{step1} is configured to 8 as the first three Rank1 blocks have a total of 8 nonzero values. Furthermore, if there are enough data words stored in VMFU for the next processing step, the GLB fetch is not performed (\eg, at step 2 in Fig.~\ref{fig:sparseB-vmfu-23}, VFMU has 13 valid entries, and the next processing step only needs 8, so no GLB fetch is performed at the step). Such a mechanism allows the metadata information to catch up with the fetched nonzeros from GLB.

The gating SAF is applied to Operand B's Rank0 sparsity to save energy by letting the \emph{MAC} unit in each PE stay idle when there is no effectual operation to perform. Note that since the gating SAF does not change the number of cycles spent, it still keeps the PEs in sync, and thus the partial sum accumulation inside each PE and across PEs is not impacted by the support for sparse Operand B.


\section{Experimental Results}
\label{sec:results}
In this section, we discuss our experimental setup and present results related to the co-designed hardware (\ie, \designName) and the software (\ie, HSS-based sparsification).

\subsection{Methodology}
\label{sec:methodology}


\subsubsection{Baseline designs}
Given the abundant prior designs, we compare \designName \ to state-of-the-art representative designs that capture the key properties of each category described in Table~\ref{tab:hss-existing-work}. 
\begin{itemize}[topsep=0pt,itemsep=0pt,partopsep=0pt,parsep=0pt, leftmargin=12pt]
    \item TC~\cite{volta-white-paper} represents dense accelerators (\eg, ~\cite{volta-white-paper, tpu-datacenter, diannao}), which are oblivious of the potential benefits introduced by sparsity.
    \item STC~\cite{ampere-white-paper} represents single-sided G:H structured sparse accelerators (\eg, \cite{ampere-white-paper, vector-sparse-tensor-core}), which introduce considerable efficiency benefits at a reasonable sparsity tax, but often support a limited number of sparsity degrees.
    \item S2TA~\cite{s2ta} represents dual-sided (\ie, both operands) G:H structured sparse accelerators, which improve on the single-sided designs with additional efficiency gains from the second sparse operand, but can introduce higher sparsity tax.
    \item  DSTC~\cite{dstc} represents dual-sided unstructured sparse accelerators (\eg, \cite{dstc, scnn, eyeriss-v2, gamma, sparten}), which often have high flexibility but introduce considerable sparsity tax.
\end{itemize}
Table~\ref{tab:sparsity-support} describes more details on each design's supported sparsity structures (if any) for each operand. 

To ensure fairness, as shown in Table~\ref{tab:exp-setup}, we allocate similar storage and compute resources to all designs. For designs that support compression of sparse input activations (\ie, DSTC, S2TA, and \designName), the same-style compression unit support as shown in Fig.~\ref{fig:hierarchical-safs} is applied to reduce DRAM traffic. Furthermore, all accelerators are designs that process DNNs as matrix multiplications. 
Since matrix multiplications accelerators treat the two operands interchangeably, we allow them to swap operands and report the best hardware performance (\eg, since STC benefits from sparse operand A, we swap the operands if operand B is sparse and A is dense). 

\renewcommand{\arraystretch}{1.2}
\begin{table}[tb]
\centering
\resizebox{\columnwidth}{!}{
\begin{tabular}{c|cc}
\multirow{2}{*}{\textbf{Design}} & \multicolumn{2}{c}{\textbf{Supported Sparsity Patterns}}                                                                                                                                                                    \\ \cline{2-3} 
                                 & \multicolumn{1}{c|}{\textbf{Operand A}}                                                                                          & \textbf{Operand B}                                                                          \\ \hline
TC~\cite{volta-white-paper}                               & \multicolumn{2}{c}{dense}                                                                                                                                                                                                       \\ \hline
STC~\cite{ampere-white-paper}                              & \multicolumn{1}{c|}{dense; C$\textsubscript{0}$$($\{G$\le$2\}:4$)$}                                                                                 & dense                                                                                       \\ \hline
DSTC~\cite{dstc}                             & \multicolumn{2}{c}{dense;  unstructured sparse}                                                                                                                                                                                \\ \hline
S2TA~\cite{s2ta}                             & \multicolumn{1}{c|}{C$\textsubscript{0}$$($\{G$\le$4\}:8$)$}                                                                                       & C$\textsubscript{0}$$($\{G$\le$8\}:8$)$                                                                       \\ \hline
\begin{tabular}[c]{@{}c@{}}\designName \\ (our work)\end{tabular}                     & \multicolumn{1}{c|}{\begin{tabular}[c]{@{}c@{}} C$\textsubscript{1}$$($4:\{4$\le$ H$\le$8\}$)$$\rightarrow$C$\textsubscript{0}$$($2:\{2$\le$ H$\le$4\}$)$\end{tabular}} & \begin{tabular}[c]{@{}c@{}}dense; \\ unstructured sparse \end{tabular} \\ 
\end{tabular}
}
\caption{\textbf{Supported sparsity patterns for each design.}}
\label{tab:sparsity-support}
\end{table}
\renewcommand{\arraystretch}{1}

\renewcommand{\arraystretch}{1.2}
\begin{table}[tb]
\centering
\resizebox{1\columnwidth}{!}{
\begin{tabular}{c|c|c|c}
\multirow{2}{*}{\textbf{Design}}  & \multicolumn{2}{c|}{\textbf{Storage}}                               & \multirow{2}{*}{\textbf{Compute}}    \\ \cline{2-3} 
                                  & \multicolumn{1}{c|}{\textbf{GLB}} & \textbf{RF}                     &    \\ \hline

TC~\cite{volta-white-paper}      & \multicolumn{1}{c|}{320KB}        & \multirow{3}{*}{$4 \times 2$ KB} & \multicolumn{1}{c}{\multirow{3}{*}{$4 \times 256$}}  \\ \cline{1-1} \cline{2-2}

STC~\cite{ampere-white-paper}    & \multicolumn{1}{c|}{$256 + 64$KB}   &          & \multicolumn{1}{c}{}       \\ \cline{1-1} \cline{2-2}

DSTC~\cite{dstc}                 & \multicolumn{1}{c|}{$256 + 64$KB}   &          & \multicolumn{1}{c}{}       \\ \cline{1-1} \cline{2-4} 

S2TA~\cite{s2ta}                 & \multicolumn{1}{c|}{$256 + 64$KB}   & $64 \times 64$B                & \multicolumn{1}{c}{$64 \times 16$}      \\ \cline{1-1} \cline{2-4} 
HighLight (our work)             & \multicolumn{1}{c|}{$256 + 64$KB}   & $4 \times 2$KB                  & \multicolumn{1}{c}{$4 \times 256$}     \\ 
\end{tabular}
}
\vspace{2pt}
\caption{\textbf{Hardware resource allocation. GLB is partitioned to data and metadata storage for sparse designs.}}
\label{tab:exp-setup}
\end{table}
\renewcommand{\arraystretch}{1} 

\subsubsection{Workloads} 

We evaluate two classes of workloads: \textbf{(i) Synthetic matrix multiplication} with operand A and B matrices that are 1024-by-1024, a common shape in DNN workloads. A and B are of various sparsity degrees: three different degrees for A: 0\%, 50\%, 75\%, and four different degrees for B: 0\%, 25\%, 50\%, 75\%. Synthetic workloads allow us to capture the diverse sparsity characteristics in the DNN design space. \textbf{(ii) Representative DNN models} with distinct network architectures and target applications: the convolutional ResNet50~\cite{resnet} and attention-based Deit-small~\cite{deit} for image classification trained on ImageNet~\cite{imagenet} and the attention-based Transformer-Big~\cite{attention} for language translation trained on WMT16 EN-DE~\cite{wmt16}. Actual DNN models allow us to take both accuracy and hardware performance impact into the picture.

\subsubsection{Evaluation Frameworks}
\label{sec:eval_platforms}
\textbf{Accelerator Modeling:} we use the Sparseloop-Accelergy infrastructure~\cite{micro-2022-sparseloop, accelergy-iccad} to model the accelerators. Sparseloop captures each accelerator's cycle counts and component runtime activities. We added a new density model to Sparseloop to capture the characteristics of HSS.

To characterize energy and area costs, we built 65nm Accelergy estimation plug-ins to characterize various components:
\begin{itemize}[topsep=0pt,itemsep=0pt,partopsep=0pt,parsep=0pt, leftmargin=15pt]
\item \designName's design-specific SAF implementations (\ie, muxing logic and VFMU) and datapath components (\eg, adders, multipliers): synthesized RTL.
\item Small SRAMs: SRAM compiler.
\item Large SRAMs not supported by compiler: CACTI~\cite{cacti}.
\item DRAM: propriety commercial data for LPDDR4.
\end{itemize}
All accelerator designs are evaluated with the same evaluation framework to ensure fairness.

\textbf{DNN Pruning:} We use Condensa~\cite{condensa} to introduce various sparsity patterns to various DNNs, structured or unstructured.  Since a good set of sparsity patterns should allow reasonable accuracy recovery even without novel or advanced pruning algorithms (\eg, special ways to perform hyper-parameter searches), we reuse the pruning algorithm proposed for sparse tensor core (STC)~\cite{NM-finetune}. Specifically, the algorithm for STC first statically prunes a pre-trained dense DNN by masking the appropriate weights and their gradients to zeros based on sparsity-pattern-specific sparsification rules (\eg, the HSS-based rules in Sec.~\ref{sec:hss_sparsification}), and it then fine-tunes the masked DNN to regain accuracy. 
\emph{To ensure fairness, all of our performed pruning follows the same algorithm, and the same set of hyperparameters is used for all sparsity patterns.}

\subsection{Outperforms Prior Work}
\begin{figure*}[t]%
    \centering
    \includegraphics[width=\linewidth]{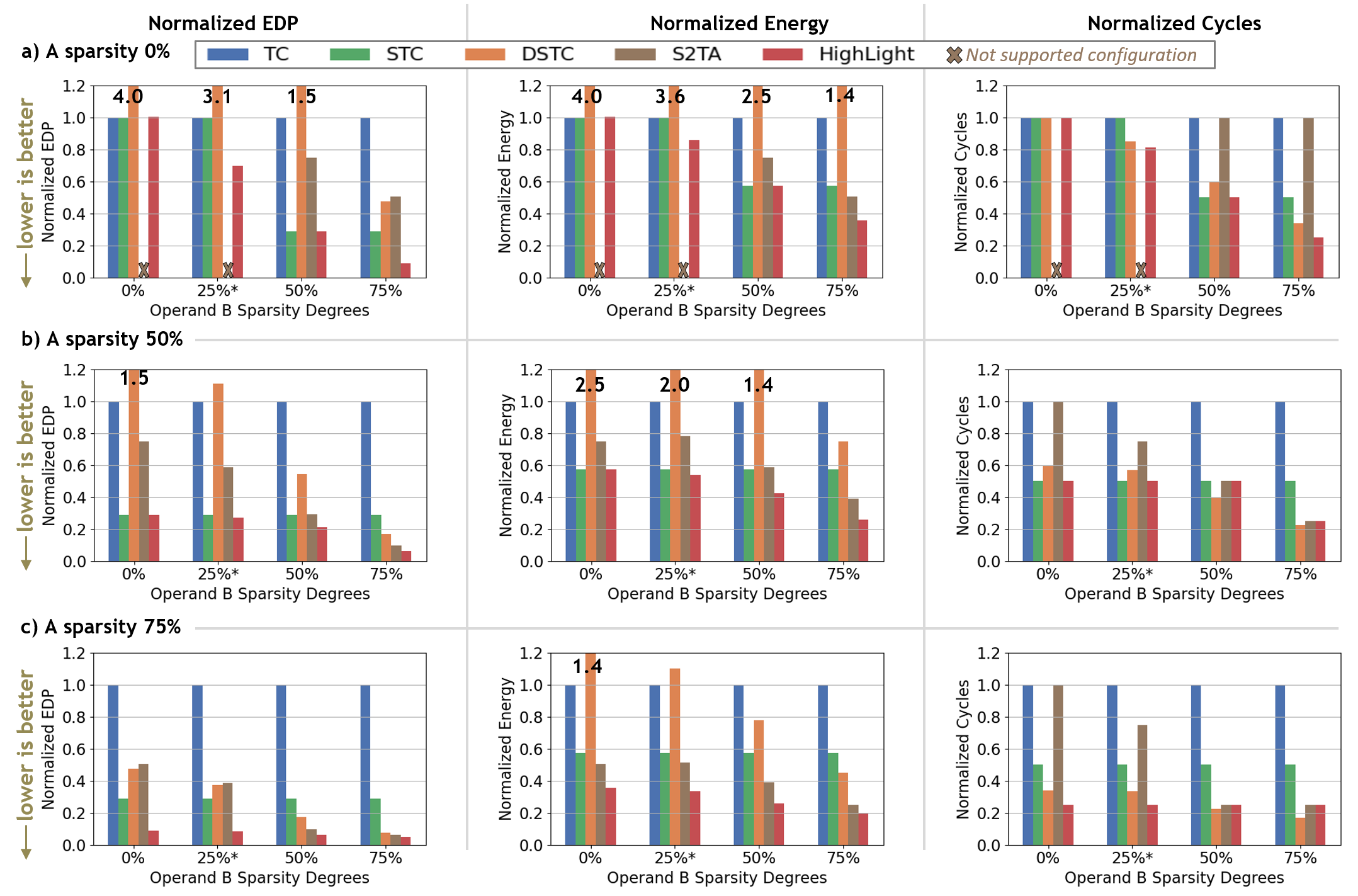} 
    \caption{ \textbf{Comparison of existing designs running workloads with operands with different sparsity degrees. We compare the overall hardware efficiency energy-delay-product (EDP), energy and speed of the designs. S2TA~\cite{s2ta} assumes both operands are structured. \designName \ is always able to effectively exploit diverse sparsity degrees. \emph{*\designName \ evaluated with 20\% sparsity for conservative estimations.}}}
    \label{fig:compare-to-existing-designs}
\end{figure*}

We compare \designName \ to existing designs running synthetic workloads to demonstrate its flexibility and efficiency. Specifically, its ability to always achieve high processing efficiency for workloads with varying sparsity degrees.


Fig.~\ref{fig:compare-to-existing-designs} compares the processing latency, energy consumption, and energy-delay product (EDP), a widely used metric for evaluating overall hardware performance in many existing works~\cite{ruby, mindmappings, multi-DNN-dataflow}. As shown in Fig.~\ref{fig:compare-to-existing-designs}, different existing designs introduce inefficient processing at different sparsity degrees. Specifically, \textbf{(i) STC} employs simple acceleration with low sparsity tax for dense and 50\% sparse workloads. However, STC's limited sparsity support fails to exploit the available opportunities for both speedup and energy for high sparsity workloads. 
\noindent\textbf{(ii) DSTC} introduces significant sparsity tax to identify effectual operations. Specifically, it employs a dataflow that requires a costly accumulation buffer that is frequently accessed. Thus, DSTC's high sparsity tax masks the sparsity-related savings for workloads with low sparsity. Furthermore, DSTC also suffers from a not perfectly balanced workload due to the unpredictable nature of unstructured sparsity, \ie, not all compute units are active.
\noindent\textbf{(iii) S2TA} requires both operands to be structured sparse and has limited flexibility on the G values supported for each operand. For example, as shown in Table~\ref{tab:exp-setup}, S2TA requires one of the operands to have \{G$\le$4\}:8, \ie, cannot have more than 50\% sparsity.  Thus, S2TA often fails to support or does not fully exploit the available speedup for workloads with operands that have low or medium sparsity. 

 \textbf{On the other hand, \designName \ is always able to efficiently exploit various sparsity degrees.} \designName's per-rank skipping SAF and low-overhead hierarchical compression format introduce brings low sparsity tax, specifically low energy overhead. Furthermore, due to the structured sparsity, \designName \ always achieves theoretical speedup with perfect workload balancing.  Thus, as shown in Fig.~\ref{fig:compare-to-existing-designs}, \designName \ always achieves the best EDP and comparable-to-best processing speed for all evaluated sparsity degrees. \designName \ achieves the best geomean for all evaluated metrics.

\begin{figure}[t]%
    \centering
    \includegraphics[width=\linewidth]{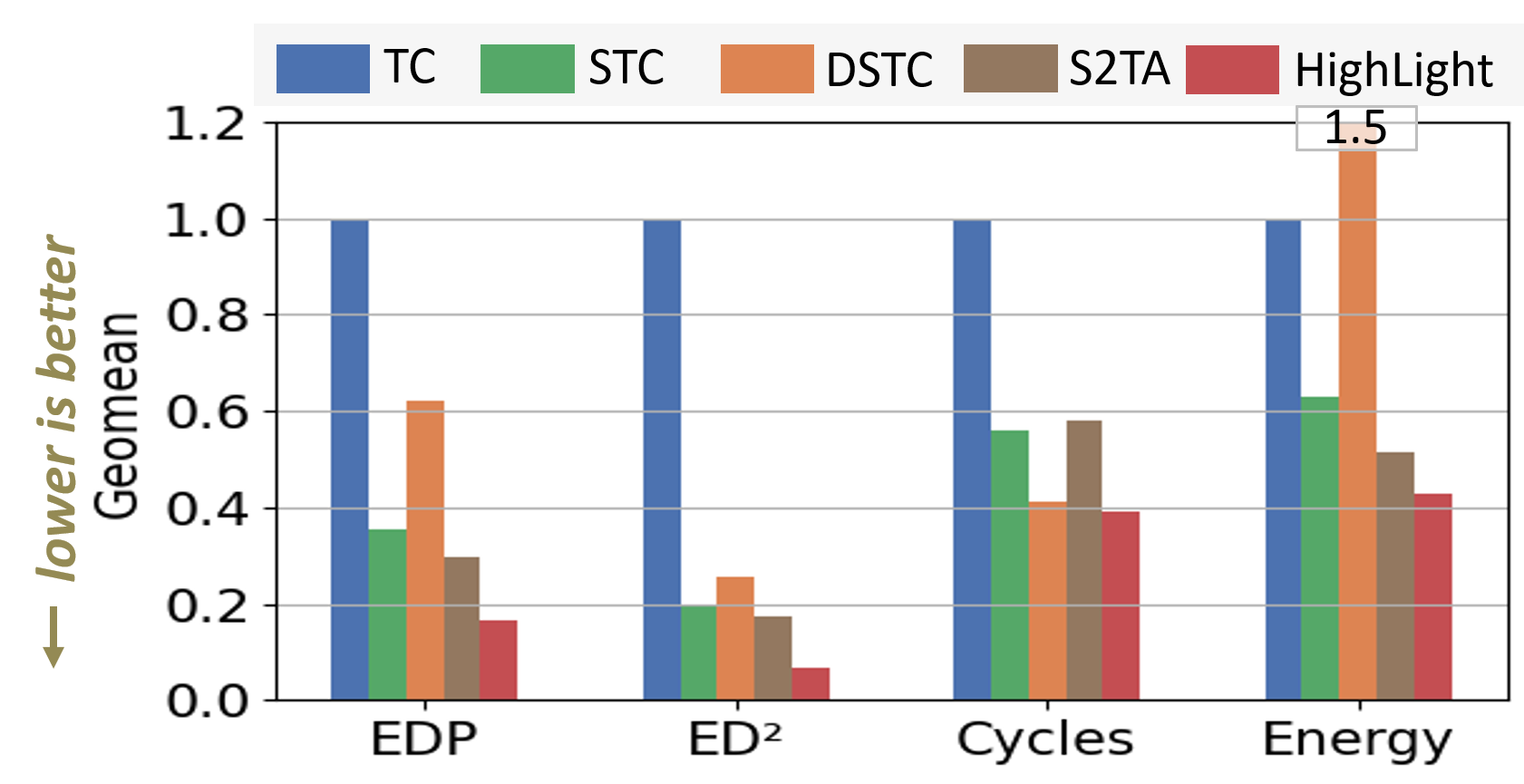} 
    \caption{ \textbf{Geomean of various metrics.  \designName \ achieves the best geomean across all evaluated metrics. }}
    \label{fig:geomean}
    \vspace{-10pt}
\end{figure}

Fig.~\ref{fig:geomean} shows each metric's geomean across the evaluated workloads. \textbf{Compared to existing designs, \designName \ achieves better geomean for all evaluated metrics.}

\subsection{Good Accuracy-Efficiency Trade-offs}
\label{sec:pareto}

\begin{figure}[tb]%
    \centering
    \includegraphics[width=\linewidth]{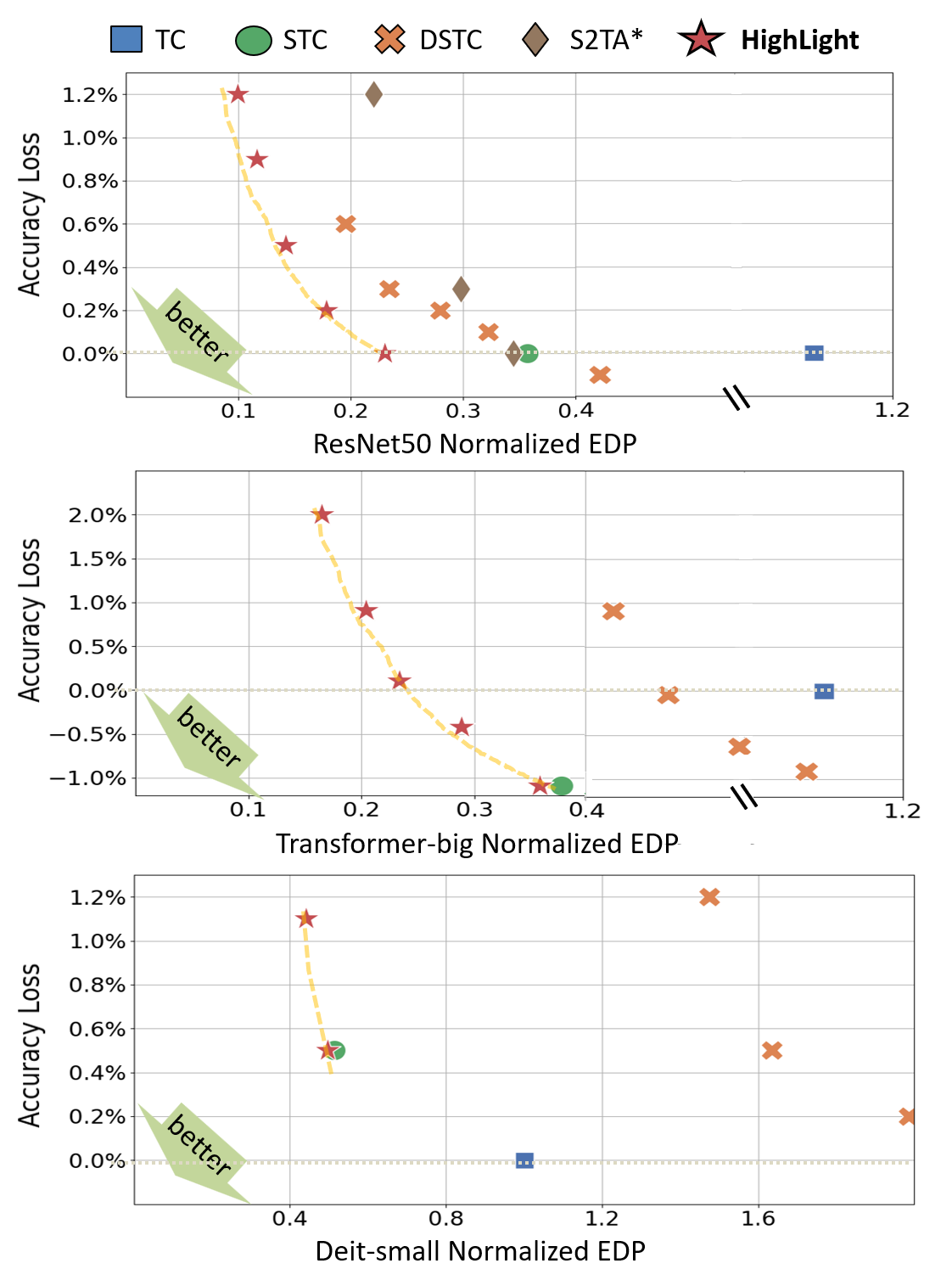} 
     \caption{ \textbf{EDP-Accuracy Loss Pareto frontier for ResNet50~\cite{resnet}, Transformer-Big~\cite{attention}, and Deit-small~\cite{deit}. Different markers refer to different accelerators.  \designName \ is always on the accuracy-EDP Pareto frontier and thus serves as a great candidate to support diverse DNNs with high hardware efficiency while maintaining accuracy.}}
    \label{fig:pareto}
    \vspace{-10pt}
\end{figure}

To demonstrate \designName \ provides good trade-offs between accuracy and efficiency, we compare the \textbf{EDP-accuracy loss relationship} of various design approaches, as shown in Fig.~\ref{fig:pareto}. Specifically, we compare \designName \ to multiple popular existing co-design approaches: \textbf{1)} dense (represented by the \textit{TC} data points); \textbf{2)} unstructured sparse (represented by the \textit{DSTC} data points); \textbf{3)} C$\textsubscript{0}$(G:H) sparse (represented by the \emph{STC} and \emph{S2TA} data points). 

We evaluate three representative DNNs: ResNet50~\cite{resnet}, Deit-small~\cite{deit}, and Transformer-Big~\cite{attention}. For ResNet50, we prune all convolutional and fully-connected layers. For Deit-small, we pruned the feed-forward block and the output projection weights.  For Transformer-big, we prune the feed-forward block and all projection weights. To ensure fairness, we use the same pruning algorithm as described in Sec.~\ref{sec:eval_platforms} for all of the evaluated sparsity patterns, including both structured and unstructured. 

Fig.~\ref{fig:pareto} shows the EDP-accuracy loss relationship for the three DNN models, with their weights pruned to different sparsity degrees.  Ideally, we would like to always have very low EDP and accuracy loss. Unfortunately, low EDP often requires higher sparsity and thus leads to higher accuracy loss. The best design should excel at balancing the trade-off, thus always sitting on the Pareto frontier of the EDP-accuracy loss relationship. Furthermore, the design should excel at the evaluated DNNs, which have different sparsity characteristics. Specifically, ResNet50 has much sparser activations than Transformer-big and Deit-small, and Deit-small has much fewer layers being pruned due to its already small parameter count (compared to other vision transformers). 

As shown in Fig.~\ref{fig:pareto}, \designName \ always sits on the Pareto frontiers. STC only delivers great accuracy-efficiency trade-off only at a single sparsity degree (\ie, 50\% sparse), S2TA fails to support attention-based models due to its incapability to process purely dense layers, and DSTC can introduce worse-than-dense EDP due to its high sparsity tax for the relatively dense models. \textbf{Thus, \designName \ serves as a great accelerator candidate to support diverse DNNs with high hardware efficiency while maintaining a reasonable accuracy loss.}  


\subsection{Sparsity Tax Evaluation}
\label{sec:tax}
\begin{figure}[t]%
    \centering
    \includegraphics[width=\linewidth]{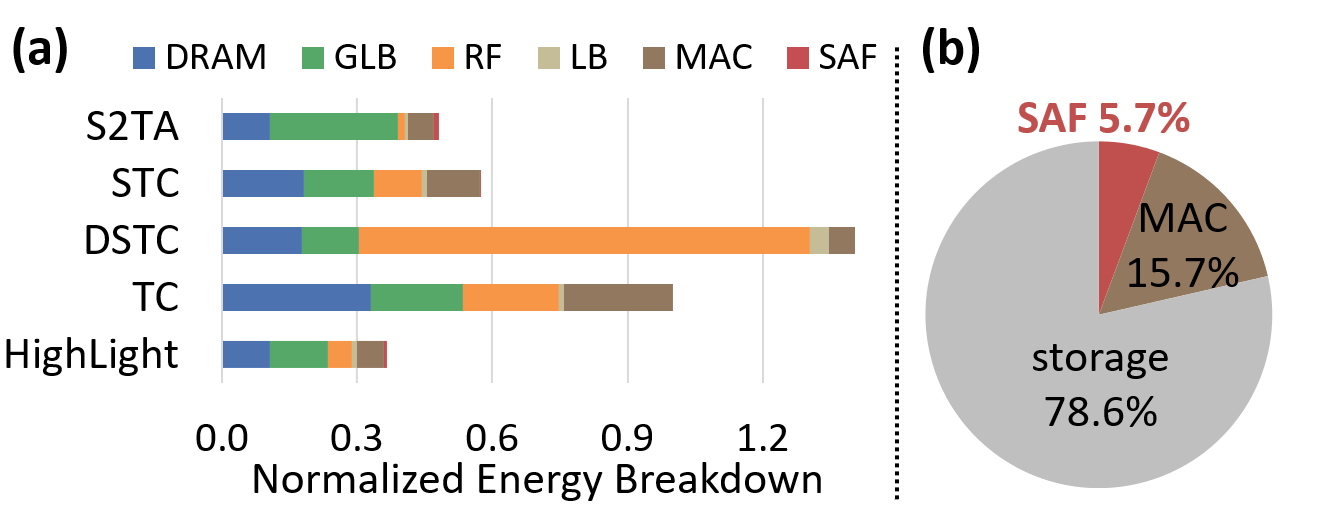} 
    \caption{ \textbf{(a) Energy consumption breakdown for a workload with 75\% sparse operand A  and dense operand B. (b) \designName \ area breakdown. \designName \ introduces low sparsity tax in terms of both energy and area.}}
    \label{fig:tax}
\end{figure}

Sparse DNN accelerators involve two types of sparsity tax: energy and area. 
Fig.~\ref{fig:tax}(a) shows the energy cost breakdown across different components in different architectures processing an example workload with 75\% operand A and a dense operand B (\ie, one example set of bars from Fig.~\ref{fig:compare-to-existing-designs}). Existing designs either do not fully exploit the sparsity for energy savings (\eg, STC only recognizes upto 50\% sparsity) or introduce inefficient dataflows to trade-off its insignificant SAF cost (\eg, DSTC suffers from significant accumulation traffic at RF due to its outer produce style dataflow). 
Fig.~\ref{fig:tax}(b) shows the area breakdown of \designName, with the SAFs accounting for only 5.7\% of the design's area. Note that since sparsity tax is intrinsic to the hardware design, different workloads would not change the general amount of sparsity tax introduced\footnote{The metadata costs differ based on workload sparsity, but such cost is not the dominant source of sparsity tax in the evaluated designs.}.  \textbf{Thus, \designName \ has low sparsity tax.}


\subsection{Potential Benefits of Dual-Side HSS}
\label{sec:hss_case_study}

Exploiting sparsity in both operands for speedup, \ie, dual-side speedup, is highly desirable but often requires complex intersection hardware and workload balancing techniques. To address such a challenge, we make the observation that,  with low intersection sparsity tax and easy workload balancing, hierarchical structured sparsity can also potentially be used to achieve dual-side speedup. In this section, we will discuss the potential benefits such improvements could bring to motivate more studies on fully supporting dual-side HSS workloads.

An HSS-based accelerator can achieve dual-side speedup with easy workload balancing by supporting multi-rank HSS operands with alternating dense ranks, \eg, weights with C$\textsubscript{1}$(dense)$\rightarrow$C$\textsubscript{0}$(2:4) and input activations (iacts) with C$\textsubscript{1}$(2:4)$\rightarrow$ C$\textsubscript{0}$(dense). To identify the nonzero value locations, each operand only needs to carry the offset metadata for the rank with G:H sparsity. For example, weights with C$\textsubscript{1}$(dense)$\rightarrow$C$\textsubscript{0}$(2:4) carry offset metadata for each nonzero value to identify its relative position in its C$\textsubscript{0}$ block, and iacts with C$\textsubscript{1}$(2:4)$\rightarrow$ C$\textsubscript{0}$(dense) carry offset metadata for each C$\textsubscript{0}$ block to identify its relative position in each C$\textsubscript{1}$ block. 

With the alternating dense ranks,  both operands are never sparse at the same rank; therefore, the SAF at each rank only needs to perform dense-sparse intersections. For instance, for C$\textsubscript{1}$, skipping can be performed by intersecting C$\textsubscript{1}$(2:4) in iacts with C$\textsubscript{1}$(dense) in weights. For C$\textsubscript{0}$, skipping can be performed by intersecting C$\textsubscript{0}$(2:4) in weights with C$\textsubscript{0}$(dense) in iacts.  Such dense-sparse intersections by nature lead to a perfectly balanced workload~\cite{GoSPA}. We refer to the accelerator design that supports dual-side HSS as the \emph{dual structured sparse operands (DSSO)} design.

\begin{figure}[t]%
    \centering
    \includegraphics[width=\linewidth]{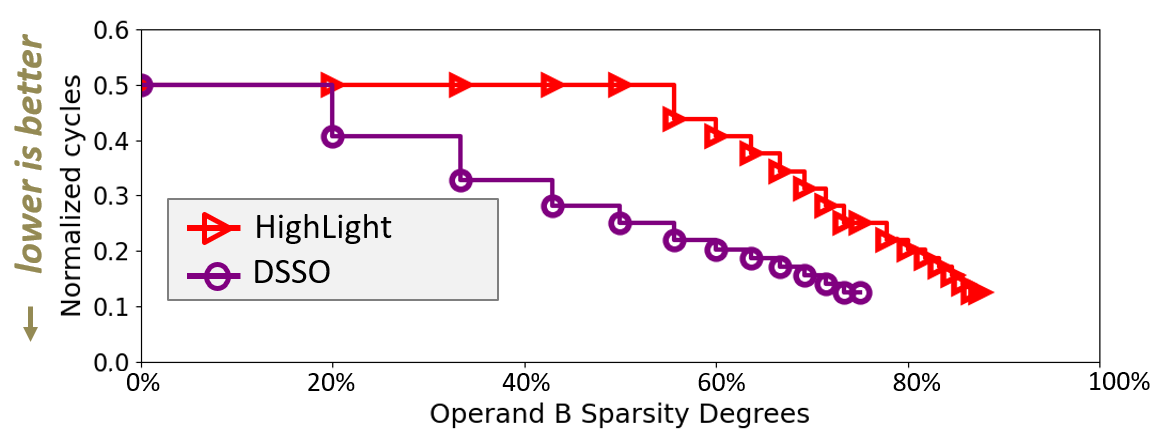} 
    \vspace{1pt}
    \caption{\textbf{Normalized processing speed of \designName \ and the dual structured sparse operand (\emph{DSSO}) design. The \emph{DSSO} design supports dual-side HSS with alternating dense ranks and allows dual-side speedup.}} 
    \label{fig:case_study}
    \vspace{-10pt}
\end{figure}

Fig.~\ref{fig:case_study} compares the processing speed of  \designName \ and \emph{DSSO} for workloads with operand A (weights) with C$\textsubscript{1}$(dense)$\rightarrow$C$\textsubscript{0}$(2:4) and operand B that follows C$\textsubscript{1}$(2:{2$\le$H$\le$8})$\rightarrow$C$\textsubscript{0}$(dense).  \emph{DSSO} demonstrates interesting trade-offs compared to singled-sided HSS. As shown in Fig.~\ref{fig:case_study}, \textbf{\emph{DSSO} achieves 2$\times$ better processing speed compared to \designName \ for the commonly supported sparsity degrees.} However, since \emph{DSSO} requires one rank to be dense to enable perfect workload balancing, there are fewer sparsity degrees supported for operand B. 

As currently proposed, the HighLight design does not naively support the dual-side HSS configuration. Specifically, HighLight does not discuss the hardware support needed to prune and compress the output activations of each layer into the desired HSS format. In addition, although existing works have shown that it is possible for DNNs with dual structured sparse operands to maintain accuracy~\cite{s2ta, DASNet-iact-pruning, NM_iact_pruning}, such DNNs may still require more advanced pruning techniques to recover accuracy. Thus, there remain many interesting research questions to answer regarding complete dual-side HSS support in both hardware and pruning algorithms.

\section{Related Work}

There is ample prior work in designing accelerators for efficiently processing sparse DNNs. These works either focus on co-designing sparsity patterns and hardware or solely focus on hardware for existing pruned models. Co-design approaches involve pruning DNNs to structured sparsity patterns that can be easily exploited by the underlying hardware. The target underlying system can be existing dense systems (maybe with relatively minor ISA updates)~\cite{patdnn, channel-pruning, gpu-aware-pruning, column-vector, NM-training}, \eg, GPUs, or custom accelerators~\cite{ampere-white-paper, vector-sparse-tensor-core, s2ta, column-combine,sdp-pim} designed for the sparsity structure. The accelerators can be designed either with conventional digital technology or emerging technology, \eg, processing-in-memory accelerators~\cite{sdp-pim}. 
To better recover accuracy loss due to the enforced structure, some proposals have relatively relaxed structures and pre-process the pruned models into more compact structures before sending them to hardware, \eg, pack unstructured columns into compact blocks~\cite{column-combine}.  Since structured sparsity has static nonzero values locations, the accelerators often have a low sparsity tax but low flexibility. 

On the other hand, accelerators designed for existing pruned models or general sparse matrix multiplications often involve designing flexible sparsity support for unstructured sparsity~\cite{scnn, eyeriss-v2, eyeriss-v1, sparten, gamma, extensor, outerspace, sparch, sigma, sparse-reram-engine, smash}. Since supporting dynamic nonzero value locations requires extremely flexible hardware, these designs often focus on different dataflows that reduce complexity, efficient auxiliary components (\eg, fast intersection unit~\cite{extensor}) that alleviate the significant control overhead, etc. Nonetheless, such accelerators often rely on the assumption that unstructured pruning can introduce very high ($>$ 80\%) sparsity, which can cancel out the cost of high sparsity tax.

The concept of hierarchy is also used in compressed data representations~\cite{smash, bcsr_cpu, extensor}. However, this line of work often focuses on better workload partitioning to enable efficient hardware processing, instead of using the hierarchy to provide flexibility and/or modularity, which is the goal of HSS. In fact, their proposed sparsity patterns are often unstructured or one-rank structured sparse (\eg, SMASH~\cite{smash} employs two levels of bitmask to represent unstructured sparse tensors), and target HPC/Graph analytics applications, which often have much higher sparsity degrees, and thus are less sensitive to high sparsity tax than DNNs.

\section{Conclusion}
Various optimization techniques introduce DNNs with diverse sparsity degrees. The diversity challenges the assumptions made by existing DNN accelerators, which often trade flexibility for efficiency, or vice versa.
This paper addresses the importance of balancing accelerator flexibility and efficiency by proposing a novel class of DNN sparsity patterns, hierarchical structured sparsity (HSS), which leverages the multiplication of fractions to systematically represent diverse sparsity degrees. Leveraging the modularity of HSS, we developed \designName \ to achieve flexible sparsity support with low sparsity tax. 
In conjunction, we show that HSS allows DNN developers to prune DNNs to diverse degrees while maintaining desired accuracy levels.
Compared to dense accelerators,  \designName\ achieves a geomean of 6.4$\times$ (and up to 20.4$\times$) better energy-delay product (EDP) across layers with diverse sparsity degrees (including dense) and is at parity for dense DNN layers. Compared to sparse accelerators, \designName\ achieves a geomean of 2.7$\times$ (and up to 5.9$\times$) better EDP and is at parity for sparse layers.


\begin{acks}
We would like to thank the anonymous reviewers for their constructive feedback. Part of the research was done during Yannan Nellie Wu’s internship at NVIDIA Research. This research was funded by the MIT AI Hardware Program. We would also like to thank Andrew Feldman, Michael Gilbert, Tanner Andrulis, Yifan Yang, and Zi Yu (Fisher) Xue for their valuable feedback on improving the clarity of the paper.
\end{acks}

\newpage
\bibliographystyle{ACM-Reference-Format}
\bibliography{references}

\end{document}